\documentclass[12pt, a4paper]{article}
\usepackage{epsf}
\usepackage{cite}
\usepackage{graphicx,rotating}
\usepackage{comment}
\usepackage{ascmac}
\usepackage[bookmarksopen,colorlinks=true,linkcolor=light_blue,citecolor=dark_red,urlcolor=dark_red,linktoc=all]{hyperref}
\usepackage{amsmath,mathtools,amsfonts,amssymb,slashed,braket,bm,bbm,cite,xcolor,bbold,physics}
\usepackage[footnotesize]{caption}

\usepackage[height=8.85in,width=6.5in]{geometry}

\usepackage{multicol}
\definecolor{dark_red}{rgb}{0.7, 0., 0.}
\definecolor{light_pink}{rgb}{1,0.4,0.4}
\definecolor{light_blue}{rgb}{0.284602,0.317763,0.963947}
\definecolor{darkgreen}{RGB}{0, 100, 0}
\definecolor{desy_blue}{HTML}{009EE2}
\definecolor{desy_orange}{HTML}{FD8800}
\definecolor{forestgreen}{HTML}{228B22}
\definecolor{ochre}{HTML}{CCAA2B}

\makeatletter
\@addtoreset{equation}{section}

\makeatother

\begin{document}

\begin{titlepage}

\begin{center}
\small{
\hfill UMN-TH-4209/23\\
\hfill FTPI-MINN-23-03 \\
\hfill TU-1176\\
\hfill KEK-QUP-2023-0003\\
\hfill KEK-TH-2498\\
}
\vskip 0.5in

{\Large \bfseries
Zero Casimir Force in Axion Electrodynamics\\[.2em]
and the Search for a New Force\\
}

\vskip .6in

{\large
Yohei Ema$^{(a,b)}$, 
Masashi Hazumi$^{(c,d,e,f,g)}$,
Hideo Iizuka$^{(c,h)}$, \\ \vspace{2.5mm}
Kyohei Mukaida$^{(d)}$ and
Kazunori Nakayama$^{(c,i)}$
}

\vskip .3in

\begin{tabular}{ll}
	$^{(a)}$& \!\!\!\!\!\emph{William I. Fine Theoretical Physics Institute, School of Physics and Astronomy, 
	}\\[-.15em]
	& \!\!\!\!\!\emph{University of Minnesota, Minneapolis, MN 55455, USA}\\
	$^{(b)}$& \!\!\!\!\!\emph{School of Physics and Astronomy, University of Minnesota,}\\[-.15em]
	& \!\!\!\!\!\emph{Minneapolis, MN 55455, USA}\\
	$^{(c)}$& \!\!\!\!\!\emph{International Center for Quantum-field Measurement Systems}\\[-.15em]
	& \!\!\!\!\!\emph{for Studies of the Universe and Particles (QUP), KEK, Tsukuba, Ibaraki 305-0801, Japan}\\
	$^{(d)}$& \!\!\!\!\!\emph{Institute of Particle and Nuclear Studies (IPNS), KEK, Tsukuba, Ibaraki 305-0801, Japan}\\
	$^{(e)}$& \!\!\!\!\!\emph{Japan Aerospace Exploration Agency (JAXA),}\\[-.15em]
	&\!\!\!\!\!\emph{Institute of Space and Astronautical Science (ISAS), Sagamihara, Kanagawa 252-5210, Japan}\\
	$^{(f)}$& \!\!\!\!\!\emph{Kavli Institute for the Physics and Mathematics of the Universe (Kavli IPMU, WPI),}\\[-.15em]
	& \!\!\!\!\!\emph{UTIAS, The University of Tokyo, Kashiwa, Chiba 277-8583, Japan}\\
	$^{(g)}$& \!\!\!\!\!\emph{The Graduate University for Advanced Studies (SOKENDAI), Miura District,}\\[-.15em]
	& \!\!\!\!\!\emph{Kanagawa 240-0115, Hayama, Japan}\\
	$^{(h)}$& \!\!\!\!\!\emph{Toyota Central R\&D labs, Inc., Nagakute, Aichi 480-1192, Japan}\\
	$^{(i)}$& \!\!\!\!\!\emph{Department of Physics, Tohoku University, Sendai 980-8578, Japan}\\
\end{tabular}

\end{center}

\vskip .6in

\begin{abstract}
\noindent
We point out that there is a stable configuration of metal plates where the Casimir force is vanishing in axion electrodynamics. We consider a concrete setup involving Weyl semimetals, which hosts an axion-like effect on the electromagnetism, towards the measurement of the axionic effect on the Casimir force. Our setup realizes zero Casimir force between metals and may be useful for the search for new force mediated by light particles at the micrometer scale.

\end{abstract}

\end{titlepage}

\tableofcontents

\section{Introduction}

The concept of ``axion'' has been first introduced to solve the strong CP problem in quantum-chromo dynamics~\cite{Peccei:1977hh,Peccei:1977ur,Weinberg:1977ma,Wilczek:1977pj}. 
The axion is one of the well-motivated candidates of dark matter in the Universe~\cite{Preskill:1982cy,Abbott:1982af,Dine:1982ah} and there are rich cosmological phenomena related to the axon dynamics (see Refs.~\cite{Kawasaki:2013ae,Marsh:2015xka,DiLuzio:2020wdo} for reviews).
A key property of the axion, denoted by $a(x)$, is its interaction
with the photon through the anomalous coupling:
\begin{align}
	\mathcal L =  -\frac{a}{4 M} F_{\mu\nu}\widetilde F^{\mu\nu},
\end{align}
where $F_{\mu\nu}=\partial_\mu A_\nu-\partial_\nu A_\mu$ is the electromagnetic field strength tensor, $\widetilde F^{\mu\nu}=\epsilon^{\mu\nu\rho\sigma}F_{\rho\sigma}/2$ and $M$ represents the axion-photon coupling strength.
This type of hypothetical coupling is used for experimental searches of axion-like particles. Despite tremendous experimental efforts to find the axion, it is not discovered yet~\cite{Jaeckel:2010ni,Graham:2015ouw,AxionLimits}.
On the other hand, there have been significant developments in the theory of topological insulators~\cite{Kane:2005,Fu:2007,Fu:2007b,Zahid:2010,Qi:2011}. 
The electromagnetic response of topological insulators may be described by the following term in the Lagrangian:
\begin{align}
	\mathcal L =  -\frac{\theta}{4} F_{\mu\nu}\widetilde F^{\mu\nu},
\end{align}
where $\theta=\alpha_e/(4\pi)$ inside the topological insulator while it is zero outside the insulator. This provides a manifestation of the axion electrodynamics in condensed-matter systems. 
Further investigations revealed that space-time dependent $\theta(\vec x,t)$ and hence the axion-like effects may appear in the bulk of some materials.
The Weyl semimetal is a class of material whose electromagnetic response is represented by $\theta(t,\vec x)= b_\mu x^\mu = (-b_0t + \vec b\cdot \vec x)$~(see e.g., Refs.~\cite{Armitage:2017cjs,Sekine:2020ixs} for reviews). 
In this case, $\vec b$ represents a vector connecting two Weyl nodes in the electron dispersions in momentum space. 
Even a dynamical axion field may exist in condensed-matter systems~\cite{Wilczek:1987mv,Li:2010,Sekine:2020ixs} and its application to the detection of particle dark matter has been considered~\cite{Marsh:2018dlj,Schutte-Engel:2021bqm,Chigusa:2021mci}, although in this paper we consider a Weyl semimetal in which $\theta(t,\vec x)$ is not a dynamical field.

The axion electrodynamics may have significant impacts on the Casimir force. The measurement of Casimir force provides direct evidence for the quantum nature of the vacuum and the existence of the background axion-like term necessarily modifies the vacuum fluctuation.
Thus it is interesting to search for possible effects of the axion electrodynamics on the Casimir force.

The accurate measurement of Casimir forces in vacuum was first carried out~\cite{Lamoreaux:1997} in 1997, a half century after the theoretical prediction~\cite{Casimir:1948dh}. Since then, significant computational and experimental efforts have been devoted to exploring Casimir forces~\cite{Ambjorn:1981xw,Plunien:1986ca,Bordag:2001qi,Bressi:2002fr,Onofrio:2006mq,Bordag:2009zz,Klimchitskaya:2009cw,Rodriguez:2011,Woods:2016,Gong:2021,Munday:2009,Rodriguez:2010,Zhao:2019,Emig:2006,Rodriguez:2010b,Intravaia:2013,Garrett:2018,Iizuka:2019}, finding possible applications in nanomechanical systems~\cite{Chan:2001,Tang:2017}. Although Casimir forces can occur at zero temperature, it is of great importance to evaluate the temperature dependence of the Casimir forces~\cite{Lifshitz1956,dzyaloshinskii1961}, in particular, as realistic experimental setups are often at room temperature. Recent experiments include such thermal effect on Casimir forces~\cite{Sushkov:2011}. 

One of the important properties of the Casimir force is that it is attractive for dielectric bodies or conductors with a reflection symmetry~\cite{Kenneth:2006vr}. This no-go theorem is violated if one considers nonreciprocal media, such as a Weyl semimetal. 
Casimir force between Weyl semimetals has been calculated in Refs.~\cite{Wilson:2015wsa,Farias:2020qqp,Rong:2021,Li:2021kcd,Oosthuyse:2023mbs} and repulsive force has been found in some parameter regions. Casimir force between normal metals separated by a chiral medium has been calculated in Refs.~\cite{Jiang:2018,Fukushima:2019sjn,Canfora:2022xcx}, and again the repulsive Casimir force has been found. 

In this paper, we first extend the setup of Refs.~\cite{Fukushima:2019sjn} and \cite{Wilson:2015wsa}.
We include the finite-temperature effect and consider a more realistic experimental apparatus
to measure the Casimir force, by taking account of the metal plates at the outermost layers.
We prove the existence of the stationary point where the Casimir force vanishes with the use of Weyl semimetals.
We further point out that such a zero-Casimir-force setup is useful for the new force search mediated by hypothetical new light particles~\cite{Adelberger:2003zx,Adelberger:2009zz}. 
The Casimir force experiments give the most stringent constraint on the new force with the new particle mass of $\sim \mathcal O({\rm eV})$ or the distance scale of $0.1$--$1\,{\rm \mu m}$~\cite{Decca:2005qz,Decca:2007yb,Sushkov:2011md,Chen:2014oda}.
To probe the new force, the Casimir force is an obstacle, which we want to remove.\footnote{The ``Casimir-less'' experiment has been reported in Ref.~\cite{Decca:2005qz} by taking a difference between Casimir forces with different materials.}
In this sense, the zero-Casimir-force setup may provide an ideal situation for the new force search.

This paper is organized as follows.
In Sec.~\ref{sec:EM} we briefly review the properties of electromagnetic waves under the axionic effect. In particular, we derive the dispersion relation and the reflection coefficients for the system including Weyl semimetals. 
In Sec.~\ref{sec:Cas} we derive the Casimir force in several setups and show that the Casimir force can be repulsive, or even can be zero. Interestingly there is a stable point at which the Casimir force is vanishing. 
In Sec.~\ref{sec:newforce} we point out that such a Casimir-free setup may be useful for a new force search mediated by hypothetical new light particles.

\section{Electromagnetic waves in axion electrodynamics}  \label{sec:EM}

\subsection{Maxwell equations and dispersion relation}

The action for the axion electrodynamics is
\begin{align}
	S = \int d^4x\left( -\frac{1}{4}F_{\mu\nu}F^{\mu\nu} - \frac{\theta}{4} F_{\mu\nu}\widetilde F^{\mu\nu} \right),
\end{align}
where
\begin{align}
	F_{\mu\nu} = \partial_\mu A_\nu-\partial_\nu A_\mu,~~~~~~\widetilde F^{\mu\nu} = \frac{1}{2}\epsilon^{\mu\nu\rho\sigma} F_{\rho\sigma},
\end{align}
and $\theta(t,\vec x)$ denotes the axion.
The equation of motion is given as 
\begin{align}
	\Box A^\mu -\partial^\mu(\partial_\nu A^\nu)-\frac{1}{2}\epsilon^{\mu\nu\rho\sigma} \partial_\nu (\theta F_{\rho\sigma}) = 0,
\end{align}
where $\Box = -\partial_t^2 + \vec\nabla^2$. Hereafter we take the Lorentz gauge $\partial_\nu A^\nu = 0$. In this gauge, the equation of motion becomes
\begin{align}
	\Box A^\mu - \frac{1}{2}\epsilon^{\mu\nu\rho\sigma} \partial_\nu (\theta F_{\rho\sigma}) = 0.
\end{align}
Let us define
\begin{align}
	b_\mu = (b_0,\vec b) \equiv \partial_\mu \theta = (\dot\theta, \vec\nabla\theta).
\end{align}
The Maxwell equation is 
\begin{align}
	&\vec\nabla\cdot\vec E = -\vec b\cdot\vec B, \\
	&\dot{\vec E}-\vec\nabla\times\vec B=-(b_0 \vec B + \vec b\times \vec E),\\
	&\vec\nabla\times\vec E + \dot{\vec B} = 0, \\
	&\vec\nabla\cdot\vec B = 0.
\end{align}
The equation of motion of the electric field is
\begin{align}
	\ddot{\vec E} - \vec\nabla^2\vec E+(b_0 \vec B)^{\dot\,}+ (\vec b\times\vec E)^{\dot\,}-\vec\nabla(\vec b\cdot\vec B)= 0.
	\label{eomE}
\end{align}

In the following, we assume that $b_\mu$ does not depend on $\vec x$ nor $t$.
Let us derive the dispersion relation of the photon in such a case 
by substituting the following ansatz:
\begin{align}
	\vec E  = \vec\epsilon e^{i(-\omega t + \vec k\cdot\vec x)},~~~~~~
	\vec B  = \vec\eta e^{i(-\omega t + \vec k\cdot\vec x)}.
\end{align}
Note that the Maxwell equation (Bianchi identity) implies $\vec\eta=\vec k \times\vec \epsilon / \omega$. From (\ref{eomE}), we have
\begin{align}
	(-\omega^2+ \vec k^2)\vec\epsilon = i b_0 (\vec k\times \vec \epsilon) + i\omega(\vec b\times \vec\epsilon)- \frac{i\vec k}{\omega}\left(\vec k\cdot(\vec b\times \vec\epsilon)\right).  \label{eomE2}
\end{align}
By multiplying $\vec k$ to both sides of this equation, we obtain
\begin{align}
	\vec k\cdot \vec\epsilon =-\frac{i}{\omega}\vec k\cdot(\vec b\times \vec\epsilon).
\end{align}
This implies that the longitudinal polarization of $\vec E$ is not zero in the presence of $\vec b$. 
Instead, we can define the divergence-free field through
\begin{align}
	\vec k\cdot \vec D =0; ~~~~~~\vec D = \vec E + \frac{\vec b\times \vec E}{\omega}.  \label{EtoD}
\end{align}
Eq.~(\ref{eomE2}) is written as
\begin{align}
	(-\omega^2+ \vec k^2)\vec\epsilon = i b_0 (\vec k\times \vec \epsilon) + i\omega(\vec b\times \vec\epsilon)+ \vec k \left(\vec k\cdot \vec\epsilon\right).
	 \label{eomE3}
\end{align}

\subsubsection{The case of $b_0=0$}

First, we assume $b_0=0$. In the remainder of this paper, we focus on this case since 
our main focus is on Weyl semimetals described by nonzero $\vec b$.
Without loss of generality, we can take $\vec b = (0,0, b)$. Eq.~(\ref{eomE3}) is rewritten in the matrix form as
\begin{align}
	\mathcal M
	\begin{pmatrix} 
		\epsilon_x \\ \epsilon_y \\ \epsilon_z
	\end{pmatrix} = 0,~~~~~~
	\mathcal M \equiv
	\begin{pmatrix} 
		-\omega^2 + k^2-k_x^2 & -k_xk_y + i\omega b & -k_xk_z \\
		-k_xk_y - i\omega b & -\omega^2 + k^2-k_y^2 &  -k_yk_z \\
		-k_xk_z & -k_y k_z & -\omega^2 + k^2-k_z^2
	\end{pmatrix}.
	\label{Mepsilon}
\end{align}
The dispersion relation is derived from the condition $\det \mathcal M = 0$. It is calculated as
\begin{align}
	\det \mathcal M = \omega^2 \left[ (\omega^2-k^2+k_z^2) b^2 - (\omega^2-k^2)^2\right] = 0.
\end{align}
For $\omega\neq 0$, the solution is given by~\cite{Wilson:2015wsa}
\begin{align}
	\omega^2 &= \vec k^2 + \frac{b^2}{2} \pm \frac{b}{2}\sqrt{b^2 + 4k_z^2}
	\nonumber \\
	 &=k_\parallel^2 + \left( \sqrt{k_z^2+ \frac{b^2}{4}} \pm\frac{b}{2}\right)^2.
	 \label{omega2_weyl}
\end{align}
This is the dispersion relation of the photon under the axion background with constant $\vec b = \vec \nabla\theta$.
In the limit $k_\parallel = 0$ and $k_z\to 0$, 
\begin{align}
	\omega^2 \simeq \begin{cases}
		b^2 \\
		k_z^4/b^2
	\end{cases}.
\end{align}
Thus we have one gapped and one gapless mode.

\subsubsection{The case of $\vec b=0$}

Next, we consider the case of $\vec b=0$ and $b_0 = {\rm const}$.
Eq.~(\ref{eomE2}) reads
\begin{align}
	(-\omega^2+ \vec k^2)\vec\epsilon = i b_0 (\vec k\times \vec \epsilon).
\end{align}
By multiplying $\vec k$ on both sides, we immediately obtain $\vec k\cdot\vec\epsilon=0$, i.e., there is no longitudinal polarization. Thus we can take $\vec k=(0,0,k)$ and $\vec \epsilon=(\epsilon_x, \epsilon_y,0)$ without loss of generality.
Then, by defining $\epsilon_{\pm} \equiv \epsilon_x\pm i\epsilon_y$, we have
\begin{align}
	(-\omega^2 + k^2 \pm b_0 k) \epsilon_\pm = 0.
\end{align}
Thus the dispersion relation is given by
\begin{align}
	\omega^2 = k (k \pm b_0).
\end{align}
This implies that the $\epsilon_-$ mode becomes tachyonic for $k\to 0$. The phenomenological consequences of this effect are often discussed in the context of axion cosmology and astrophysics~\cite{Carroll:1989vb,Harari:1992ea}. In this paper, we do not discuss this case.


\subsection{Polarization vectors}

\subsubsection{Polarization of electromagnetic waves in the vacuum}

\begin{figure}[t]
\begin{center}
   \includegraphics[width=15cm]{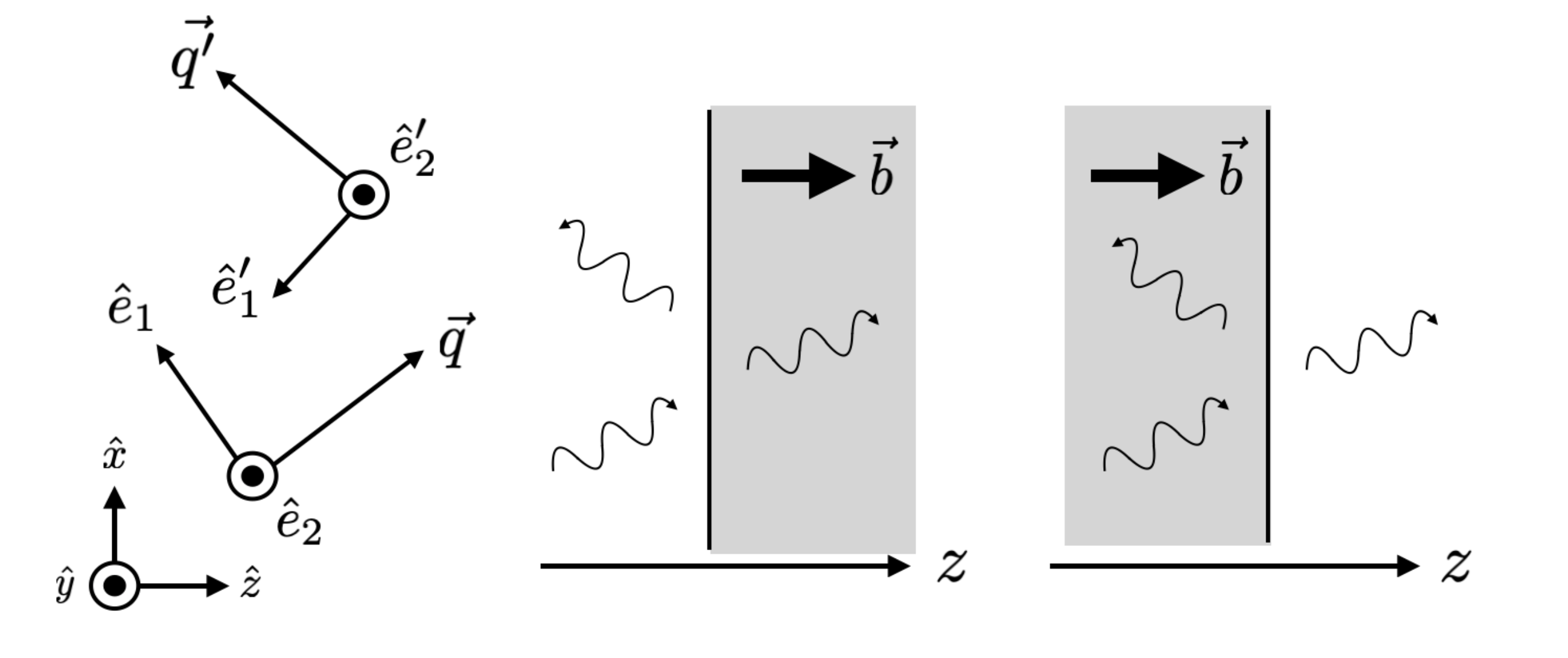}
  \end{center}
  \caption{(Left) Polarization basis of the electric field. (Middle) Injection of electromagnetic waves from vacuum to Weyl semimetal. (Right) Injection of electromagnetic waves from Weyl semimetal to vacuum. }
  \label{fig:pol}
\end{figure}

The incoming and reflected momenta are $\vec q$ and $\vec q'$, respectively (see Fig.~\ref{fig:pol}). 
Without loss of generality, we take the $\hat y$ component of the momentum to be zero by using the rotational symmetry on $(x,y)$-plane.
The incoming and reflected momenta are related as
\begin{align}
	\vec q = (q_x,0,q_z),~~~~~~\vec {q'} = (q_x, 0, -q_z).
\end{align}
We also define $q\equiv |\vec q| = |\vec {q'}|$. The dispersion relation implies $\sqrt{\epsilon}\omega = q$.

In the vacuum, the electric field satisfies $\vec q\cdot \vec E = 0$. Thus we can define the polarization of electric fields as follows (see Fig.~\ref{fig:pol}): 
\begin{align}
	&\hat e_{2} = \hat y,~~~~~~\hat e_{1} = \hat y\times \hat q = \frac{q_z}{q}\hat x- \frac{q_x}{q}\hat z; \\
	&\hat e'_{2} = \hat y,~~~~~~\hat e'_{1} = \hat y\times \hat {q'} = -\frac{q_z}{q}\hat x- \frac{q_x}{q}\hat z.
\end{align}
The $\hat e_{2}$ and $\hat e'_{2}$ polarizations correspond to the TE mode, while $\hat e_{1}$ and $\hat e'_{1}$ polarization correspond to the TM mode.
The right and left chiral modes are defined as
\begin{align}
	& \hat e_{R,L} = \frac{1}{\sqrt 2}\left( \hat e_{1} \pm i \hat e_{2} \right)  =  \frac{1}{\sqrt 2}\left( \frac{q_z}{q}\hat x \pm i \hat y -\frac{q_x}{q}\hat z\right),  \label{e_RL}\\
	& \hat e'_{R,L} =  \frac{1}{\sqrt 2}\left( \hat e'_{1} \pm i \hat e'_{2} \right)  =  \frac{1}{\sqrt 2}\left(-\frac{q_z}{q}\hat x \pm i \hat y -\frac{q_x}{q}\hat z\right).  \label{e'_RL}
\end{align}

The magnetic field is given by $\vec B = \vec q \times \vec E/\omega$. Thus the polarization vector for the magnetic field is
\begin{align}
	& \hat e^{(B)}_{R,L} = \hat q\times\hat e_{R,L} =  \frac{1}{\sqrt 2}\left(\mp\frac{iq_z}{q}\hat x + \hat y \pm \frac{iq_x}{q}\hat z\right),\\
	& \hat e^{'(B)}_{R,L} = \hat {q'}\times\hat e'_{R,L} =  \frac{1}{\sqrt 2}\left(\pm\frac{iq_z}{q}\hat x + \hat y \pm \frac{iq_x}{q}\hat z\right).
\end{align}

\subsubsection{Polarization of electromagnetic waves in the Weyl semimetal}

The dispersion relation in the Weyl semimetal is given by (\ref{omega2_weyl}).
For given $\omega$, there are two solutions for $k_z$:
\begin{align}
	(k_z^{\pm} )^2 = \kappa_z (\kappa_z\pm b).
\end{align}
where $\kappa_z \equiv \sqrt{\omega^2- k_x^2}$. 

Note that $\vec E$ in the Weyl semimetal is not transverse: $\vec k\cdot \vec E \neq 0$ (see Eq.~(\ref{EtoD})).
From (\ref{Mepsilon}), by taking $k_y=0$, the electric field in the Weyl semimetal satisfies
\begin{align}
	\mathcal M
	\begin{pmatrix} 
		\epsilon_x \\ \epsilon_y \\ \epsilon_z
	\end{pmatrix} = 0,~~~~~~
	\mathcal M \equiv
	\begin{pmatrix} 
		-\omega^2 + k^2-k_x^2 & i\omega b & -k_xk_z \\
		- i\omega b & -\omega^2 + k^2 &  0 \\
		-k_xk_z & 0 & -\omega^2 + k^2-k_z^2
	\end{pmatrix}.
\end{align}
By solving this, we obtain the polarization vector for the $\pm$ mode as
\begin{align}
	\hat e_{\pm} = \frac{1}{\sqrt 2 N_\pm}\left( \kappa_z^2 \hat x \pm i\omega \kappa_z \hat y-k_x k_z^\pm \hat z\right),
\end{align}
where $N_\pm$ is the normalization constant to make $\hat e_\pm$ the unit vector. Similarly, polarization vector for the left-moving mode $\vec k' = (k_x,0,-k_z^\pm)$ is
\begin{align}
	\hat e'_{\pm} = \frac{1}{\sqrt 2 N_\pm}\left( -\kappa_z^2 \hat x \mp i\omega \kappa_z \hat y-k_x k_z^\pm \hat z\right).
\end{align}
Note that $\hat e_{+}$ and $\hat e'_{-}$ are right-handed waves and $\hat e_{-}$ and $\hat e'_{+}$ are left-handed waves, as can be seen by comparing them in the $b\to 0$ limit with (\ref{e_RL}) and (\ref{e'_RL}).

For the magnetic field, the polarization vector is given by
\begin{align}
	\hat e^{(B)}_{\pm} = \hat k_{\pm}\times \hat e_{\pm} =  \frac{\omega}{\sqrt 2 N_\pm k_\pm}\left( \mp i\kappa_z k_z^\pm \hat x + \omega k_z^\pm \hat y \pm ik_x \kappa_z \hat z\right).
\end{align}
For reference, polarization vector for the left-moving mode $\vec k' = (k_x,0,-k_z^\pm)$ is
\begin{align}
	\hat e^{'(B)}_{\pm} = \hat k_{\pm}\times \hat e_{\pm} =  \frac{\omega}{\sqrt 2 N_\pm k_\pm}\left( \mp i\kappa_z k_z^\pm \hat x + \omega k_z^\pm \hat y \mp ik_x \kappa_z \hat z\right).
\end{align}

\subsection{Reflection coefficients}

\subsubsection{Injection from vacuum to Weyl semimetal}

The arguments below follow Refs.~\cite{Wilson:2015wsa,Farias:2020qqp,Jiang:2018}.
We consider the case of incoming photon from the vacuum, which is reflected by the Weyl semimetal as schematically shown in the middle panel of Fig.~\ref{fig:pol}.
The incoming, reflected, and transmitted waves are assumed to be of the form\footnote{
	To satisfy the boundary condition at the surface of the Weyl semimetal at any time, $\omega$ must be common on both sides. Also the boundary condition at any spatial points on the surface $\vec x_\parallel = (x,y)$ implies $(\vec q)_x = (\vec {q'})_x = (\vec k_+)_x = (\vec k_-)_x$. Then the dispersion relation implies $|(\vec q)_z| = |(\vec {q'})_z|$. Note also the relation $q\equiv |\vec q| = |\vec {q'}| = \omega$ and $\kappa_z=q_z$.
}
\begin{align}
	&\vec E_i(\vec x) =e^{i(\vec q\cdot \vec x - \omega t)}  \left( \mathcal E_{R} \hat e_{R} +  \mathcal E_{L} \hat e_{L}   \right) ,\\
	&\vec E_r(\vec x) =e^{i(\vec{q'}\cdot \vec x - \omega t)} \left( \mathcal E'_{R} \hat e'_{R} + \mathcal E'_{L} \hat e'_{L}  \right),\\
	&\vec E_\pm(\vec x) =e^{i(\vec k_+\cdot \vec x - \omega t)}\mathcal E_{+} \hat e_{+} 
	+ e^{i(\vec k_-\cdot \vec x - \omega t)}\mathcal E_{-} \hat e_{-}.
\end{align}
The corresponding magnetic fields are given by $\vec B_i = \frac{q_i\times \vec E_i}{\omega}$ and so on. As we have explained, the electric field in the Weyl semimetal is not divergence-free: $\vec k_{\pm}\cdot \vec E_{\pm} \neq 0$ and hence the TM, TE decomposition is not very useful. Instead, working on the $(L,R)$ basis will turn out to be convenient.

Let us first consider the case of incoming right-handed electromagnetic waves, i.e., $\mathcal E_{L}=0$. 
We impose continuous $\vec E_\parallel$ and $\vec B_\parallel$ at the boundary $z=0$. From $\vec E_\parallel$, we obtain
\begin{align}
	 \mathcal E'_{R} = -\frac{\omega}{N_-} q_z \mathcal E_{-},~~~~~~
	 \mathcal E_{R} - \mathcal E'_{L} =  \frac{\omega}{N_+} q_z \mathcal E_{+}.
\end{align}
From $\vec B_\parallel$, we obtain
\begin{align}
	 \mathcal E'_{R} = \frac{\omega}{N_-} k_z^- \mathcal E_{-},~~~~~~
	 \mathcal E_{R} + \mathcal E'_{L} =  \frac{\omega}{N_+} k_z^+ \mathcal E_{+}.
\end{align}
From these equations we obtain
\begin{align}
	 &\mathcal E'_{R} = \mathcal E_{-} = 0, \\
	 &\displaystyle R_{+} \equiv \frac{ \mathcal E'_{L}}{ \mathcal E_{R}} = \frac{k_z^+-q_z}{k_z^+ + q_z}=\frac{b + 2(q_z-k_z^+)}{b}.
	 \label{Rplus}
\end{align}
Therefore, the reflection of the right-handed wave on the Weyl semimetal leads to the left-handed wave. Thus it is convenient to work on the $(L,R)$ basis in the presence of Weyl semimetal.

Similarly, for the case of incoming left-handed waves, i.e., $\mathcal E_{R}=0$, we obtain from $\vec E_\parallel$
\begin{align}
	 \mathcal E'_{L} = -\frac{\omega}{N_+} q_z \mathcal E_{+},~~~~~~
	 \mathcal E_{L} - \mathcal E'_{R} =  \frac{\omega}{N_-} q_z \mathcal E_{-}.
\end{align}
and from $\vec B_\parallel$, we obtain
\begin{align}
	 \mathcal E'_{L} = \frac{\omega}{N_+} k_z^+ \mathcal E_{+},~~~~~~
	 \mathcal E_{L} + \mathcal E'_{R} =  \frac{\omega}{N_-} k_z^- \mathcal E_{-}.
\end{align}
Thus
\begin{align}
	 &\mathcal E'_{L} = \mathcal E_{+} = 0, \\
	 &\displaystyle R_{-} \equiv \frac{ \mathcal E'_{R}}{ \mathcal E_{L}} = \frac{k_z^--q_z}{k_z^- + q_z}= \frac{b + 2(k_z^- -q_z)}{b}.
	 \label{Rminus}
\end{align}

Combining them, the reflection matrix is expressed as
\begin{align}
	{\bf R}(q_z) = \frac{1}{b}
	\begin{pmatrix} 
		0 & b + 2(k_z^- -q_z) \\
		b + 2(q_z-k_z^+) & 0 
	\end{pmatrix}.
\end{align}

\subsubsection{Injection from Weyl semimetal to vacuum}

We consider the case of incoming photon from Weyl semimetal, which is reflected by the vacuum as schematically shown in the right panel of Fig.~\ref{fig:pol}.
\begin{align}
	&\vec E_i(\vec x) =e^{i(\vec k_+\cdot \vec x - \omega t)}\mathcal E_{+}\hat e_{+} + e^{i(\vec k_-\cdot \vec x - \omega t)}\mathcal E_{-}\hat e_{-},\\
	&\vec E_r(\vec x) =e^{i(\vec k'_+\cdot \vec x - \omega t)}\mathcal E'_{+} \hat e'_{+}+ e^{i(\vec k'_-\cdot \vec x - \omega t)}\mathcal E'_{-} \hat e'_{-},\\
	&\vec E_t(\vec x) =e^{i(\vec q \cdot \vec x - \omega t)} \left (\mathcal E_{R} \hat e_{R} + \mathcal E_{L} \hat e_{L}\right).
\end{align}
First let us consider the incoming $+$ mode, i.e., $\mathcal E_{-} = 0$.
In a similar way as the previous subsection, from the boundary condition, we find
\begin{align}
	& \mathcal E_{L}  = \mathcal E_{-}' = 0, \\
	& \frac{\mathcal E'_{+}}{\mathcal E_{+}} = \frac{q_z- k_z^+}{q_z + k_z^+} = -R_{+}.
\end{align}
Therefore, for the incoming plus mode, the reflected mode is also plus. As noted earlier, the incoming plus mode is (roughly) right-polarized while the reflected plus mode is left-polarized, as naturally expected. 
For the case of  incoming $-$ mode, i.e., $\mathcal E_{+} = 0$, we obtain
\begin{align}
	& \mathcal E_{R}  = \mathcal E_{+}' = 0, \\
	&\frac{\mathcal E'_{-}}{\mathcal E_{-}} = \frac{q_z- k_z^-}{q_z + k_z^-}=-R_{-}.
\end{align}

\section{Casimir force in axion electrodynamic systems}  \label{sec:Cas}

In this section, we calculate the Casimir force in axion electrodynamic systems including chiral medium or Weyl semimetals. We give two examples that exhibit the repulsive Casimir force.
A formalism for calculating the Casimir force is summarized in App.~\ref{app:Lif}. Given a setup, we can derive the function $f_\lambda(\omega)$ in (\ref{omegaflambda}) such that solutions to the equation $f_\lambda(\omega)=0$ give possible values of $\omega$. Once we find $f_\lambda(\omega)$, the finite-temperature Casimir force is evaluated by the Lifshitz formula (\ref{CasLif}). Below we consider several setups.

\subsection{Setup 1: Force between metal plates in chiral medium}

\begin{figure}[t]
\begin{center}
   \includegraphics[width=12cm]{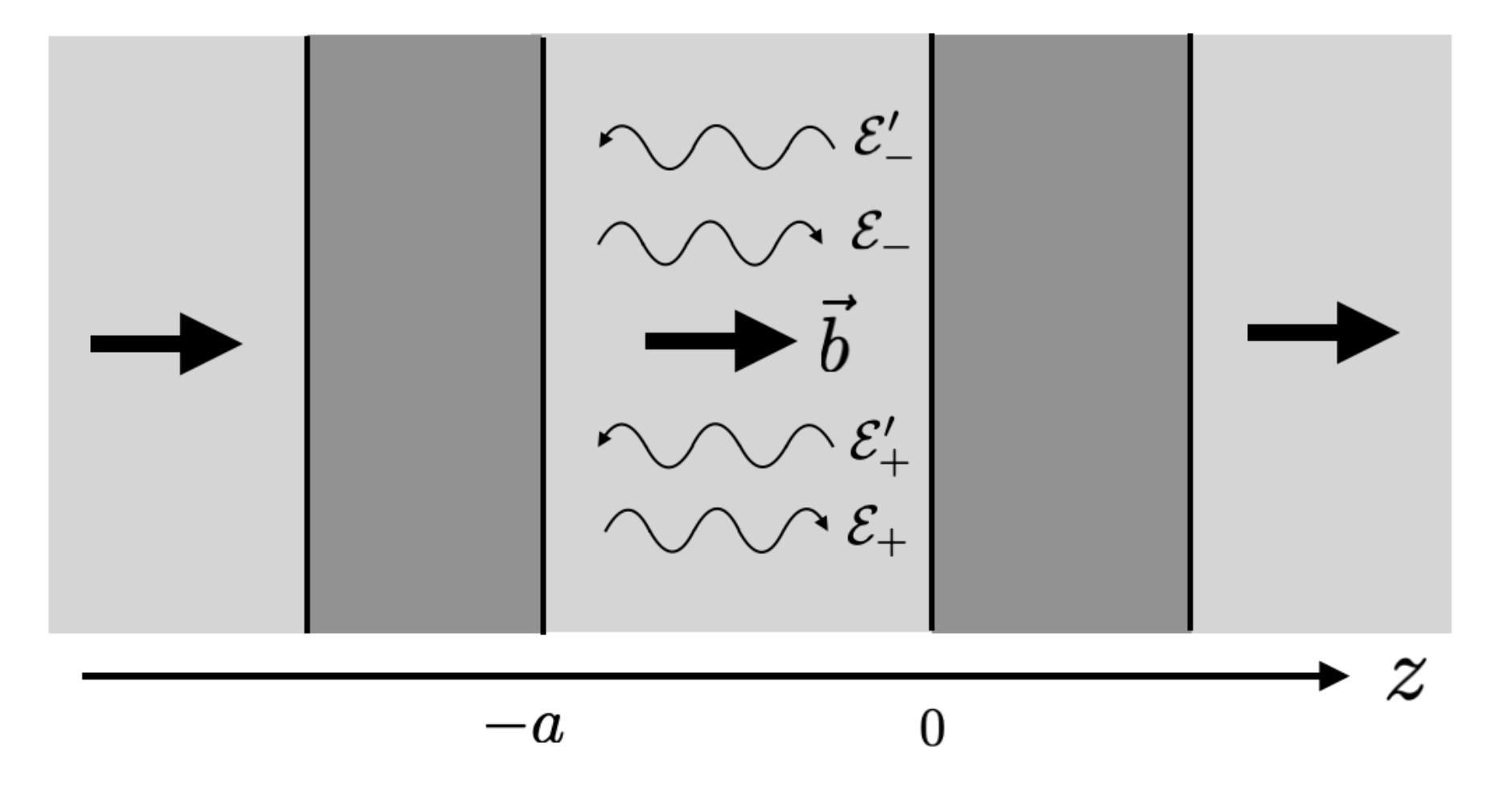}
  \end{center}
  \caption{A schematic picture of the setup 1. The chiral medium with the width $a$ is sandwiched by the perfect metal. The region outside the metal plate is also assumed to be filled with the same chiral medium. Also schematically shown are electromagnetic wave modes. The upper two and lower two modes are independent.}
  \label{fig:PWP}
\end{figure}

First let us consider the setup shown in Fig.~\ref{fig:PWP}: a Weyl semimetal sandwiched by perfect metals. Although the real Weyl semimetal is a solid and one cannot adjust its width $a$ in a given experimental setup, we can regard this system as a representative of more general setups including a chiral medium. For example, if we live at the center of a huge axionic domain wall, all the space is effectively filled by a chiral medium with constant $\vec b$. Note that the following calculations are independent of the width of the metal.\footnote{
    To be precise, a real metal has a finite skin depth and we are assuming that the width is thicker than the skin depth.
}

One solution (which we call ``+'' mode) is of the form (here and hereafter we drop the overall $e^{i(\vec k_\parallel\cdot\vec x_\parallel-\omega t)}$ factor for notational simplicity):
\begin{align}
	&\vec E = \mathcal E_+ e^{i k_z^+ z}\hat e_+ + \mathcal E'_+ e^{-i k_z^+ z} \hat e'_+, \\
	&\vec B =\frac{k_+}{\omega}\left( \mathcal E_+ e^{i k_z^+ z}\hat e_+^{(B)} + \mathcal E'_+ e^{-i k_z^+ z}\hat e^{'(B)}_+\right).
\end{align}
We impose the boundary condition $E_x=E_y=B_z=0$ at $z=0$ and $z=-a$. Then we simply find $\mathcal E_+=\mathcal E'_+$ and
\begin{align}
	1 - e^{2ik_z^+a} = 0.
\end{align}
The other solution (which we call ``$-$'' mode) is of the form
\begin{align}
	&\vec E = \mathcal E_- e^{i k_z^- z}\hat e_- + \mathcal E'_- e^{-i k_z^- z} \hat e'_-, \\
	&\vec B = \frac{k_-}{\omega}\left(\mathcal E_- e^{i k_z^- z}\hat e_-^{(B)} + \mathcal E'_- e^{-i k_z^- z}\hat e_-^{'(B)}\right).
\end{align}
From the same boundary condition, we also obtain $\mathcal E_-=\mathcal E'_-$ and 
\begin{align}
	1 - e^{2ik_z^-a} = 0.
\end{align}
We define the function $f_\lambda(\omega)$ for $\lambda = {\rm +, -}$ as
\begin{align}
	f_\lambda(\omega) =1 - e^{2ik_z^\lambda a},
\end{align}
so that the solution to $f_\lambda(\omega)=0$ gives the allowed modes. Therefore it is easily found that the solution is
\begin{align}
	k_z^+ = \frac{n\pi}{a},~~~~~~k_z^-= \frac{n\pi}{a},~~~~~~~~n=0, \pm1, \dots.
\end{align}
If $b=0$, we find $k_z^+=k_z^-=n\pi/a$ as usually found in the case of a vacuum separated by metal plates.

\begin{figure}[t]
\begin{center}
   \includegraphics[width=8cm]{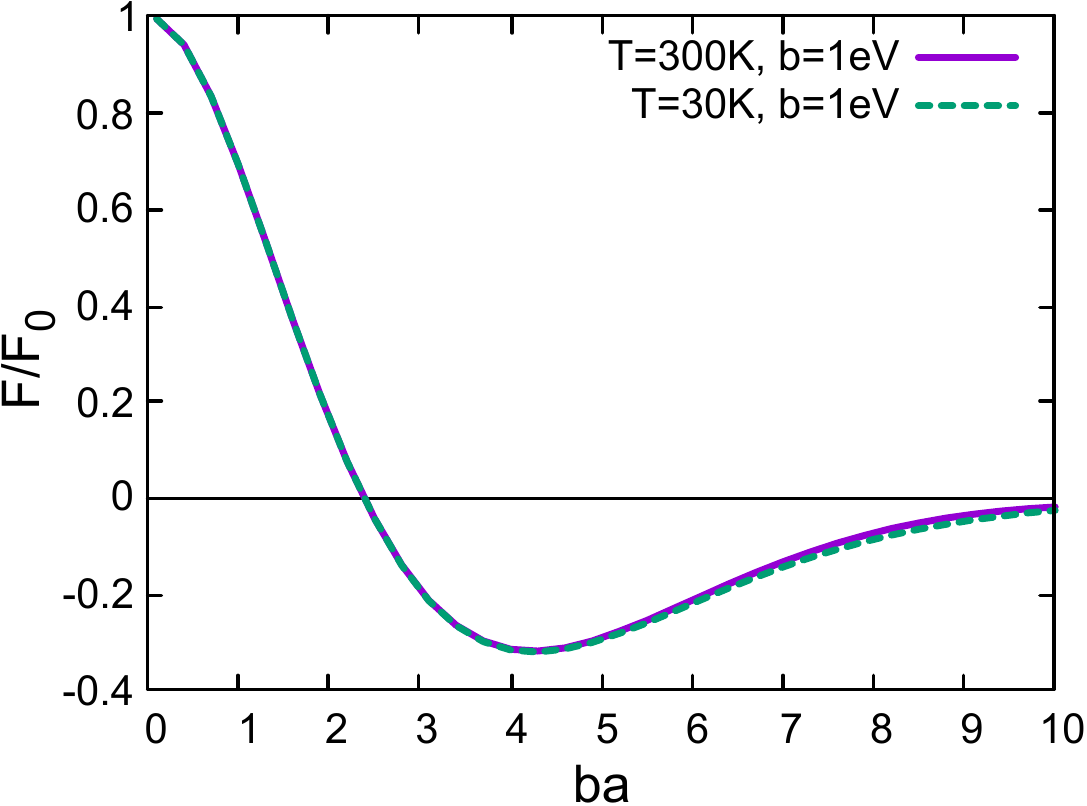}
   \includegraphics[width=8cm]{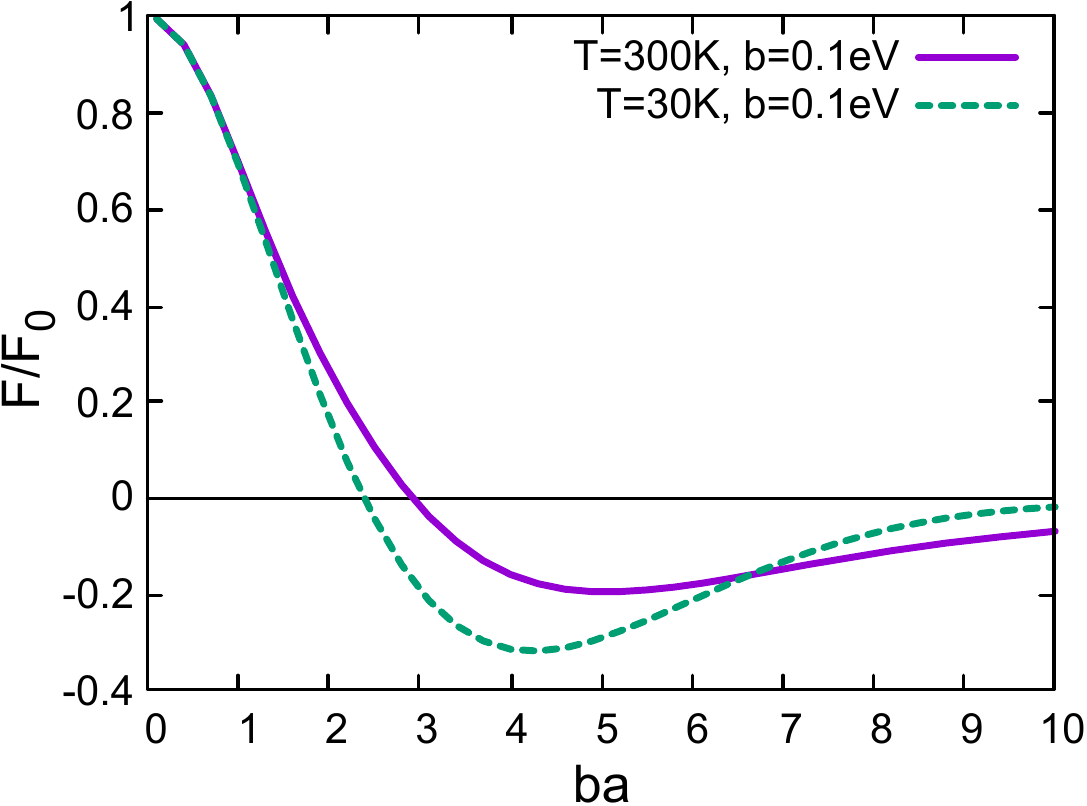}
  \end{center}
  \caption{The Casimir force normalized by the case of $b=0$ in the setup 1 (Fig.~\ref{fig:PWP}). In the left (right) panel we take $b=1\,{\rm eV}$ ($0.1\,{\rm eV}$). For reference, $1/b=0.197\,\mu {\rm m}$ in the left panel and $1/b=1.97\,\mu {\rm m}$ in the right panel. The negative $F/F_0$, below the horizontal black line, means the repulsive Casimir force. }
  \label{fig:cas_PWP}
\end{figure}

The Casimir free energy per unit area is given by
\begin{align}
	\mathcal F_{\rm Cas} &=T{\sum_{\lambda={+, -}}}{\sum_{\ell\geq 0}}'\int\frac{d^2k_\parallel}{(2\pi)^2} \ln f_\lambda(i \xi_\ell) \nonumber \\
	&=T{\sum_{\lambda={+, -}}}{\sum_{\ell\geq 0}}'\int\frac{d^2k_\parallel}{(2\pi)^2} \ln \left[1- e^{-2\tilde k_{z,\ell}^\lambda a}\right],
	\label{Fcas_setup1}
\end{align}
where $\xi_\ell = 2\pi T\ell$ and
\begin{align}
	\tilde k_{z,\ell}^{\pm} = \left[ \sqrt{\xi_\ell^2+ k_\parallel^2} \left( \sqrt{\xi_\ell^2+ k_\parallel^2} \mp ib\right) \right]^{1/2}.
\end{align}
The Casimir force is given by
\begin{align}
	F_{\rm Cas} &= -\frac{\partial \mathcal F_{\rm Cas}}{\partial a} \nonumber \\
	&=-2T\sum_{\lambda={+, -}}{\sum_{\ell\geq 0}}' \int \frac{d^2 k_\parallel}{(2\pi)^2}\frac{\tilde k_{z,\ell}^\lambda e^{-2\tilde k_{z,\ell}^\lambda a}}{1- e^{-2\tilde k_{z,\ell}^\lambda a}}.
\end{align}
Note that we are assuming that the space outside the metal is also filled by the same chiral medium with constant $\vec b$ as shown in Fig.~\ref{fig:PWP}, although it is rather implicit and hidden in a regularization procedure in (\ref{Fcas_setup1}). Otherwise, we would have an extra contribution to the Casimir force. 
The numerical results for the Casimir force, normalized by the case of $b=0$, i.e., $F_0\equiv F(b=0)=-\pi^2/(240a^4)$, are shown in Fig.~\ref{fig:cas_PWP}. 
In the left (right) panel we take $b=1\,{\rm eV}$ ($0.1\,{\rm eV}$). For reference, $1/b=0.197\,\mu {\rm m}$ in the left panel and $1/b=1.97\,\mu {\rm m}$ in the right panel.\footnote{
	Note that our definition of $b$ is the same as $b$ in Ref.~\cite{Fukushima:2019sjn} and $e^2/(2\pi^2)$ times $b$ in Ref.~\cite{Wilson:2015wsa}. See e.g., Ref.~\cite{Kotov:2018} for reference values of $b$. In our definition, it is typically $b\sim 1\,{\rm eV}$. 
}
It reproduces the result of Ref.~\cite{Fukushima:2019sjn} in the low-temperature limit. It is seen that for $b=0.1$\,eV the result for the room temperature deviates from the result of the zero-temperature limit.
An important feature is that $F/F_0$ becomes negative for a large distance, meaning that the Casimir force becomes repulsive. 

One may understand the origin of this repulsive force as follows. 
Let us consider a one-dimensional analog of Eq.~\eqref{omega2_weyl} whose dispersion relation is $\omega_\pm = \sqrt{k_z^2 + b^2/4} \pm b/2$ with $k_z = n \pi / L$.
Its energy at zero temperature is obtained by taking a summation of $\sum_{\lambda = +,-}\sum_{n \geq 0} \omega_\lambda$ with an appropriate regularization.
For simplicity, we take a regularization scheme where the high energy modes exceeding a cutoff scale $\Lambda (\gg b)$ are damped immediately.
If one takes the summation of $\lambda = +,-$ first, the resultant dispersion is just two massive modes of $\tilde \omega = \sqrt{k_z^2 + b^2/4}$, which is known to give the attractive force.
However, these massive modes only cannot take into account all the contributions because there always exist more minus modes than plus modes fulfilling $\omega_\lambda < \Lambda$.
As a result, the total energy must involve an opposite contribution to that of massive modes only.\footnote{
	In this one-dimensional analog system, it can be estimated as $\sim - b^2 L$ for $b L \lesssim \pi$.
}
Since the contribution from the ``massive'' modes is suppressed for $bL \gtrsim 1$, this leads to the repulsive Casimir force at a large distance.
Essentially the same argument also applies to the three dimensional case.

\subsection{Setup 2: Force between Weyl semimetals in vacuum} 
\label{sec:PWVWP}

\begin{figure}[t]
\begin{center}
   \includegraphics[width=14cm]{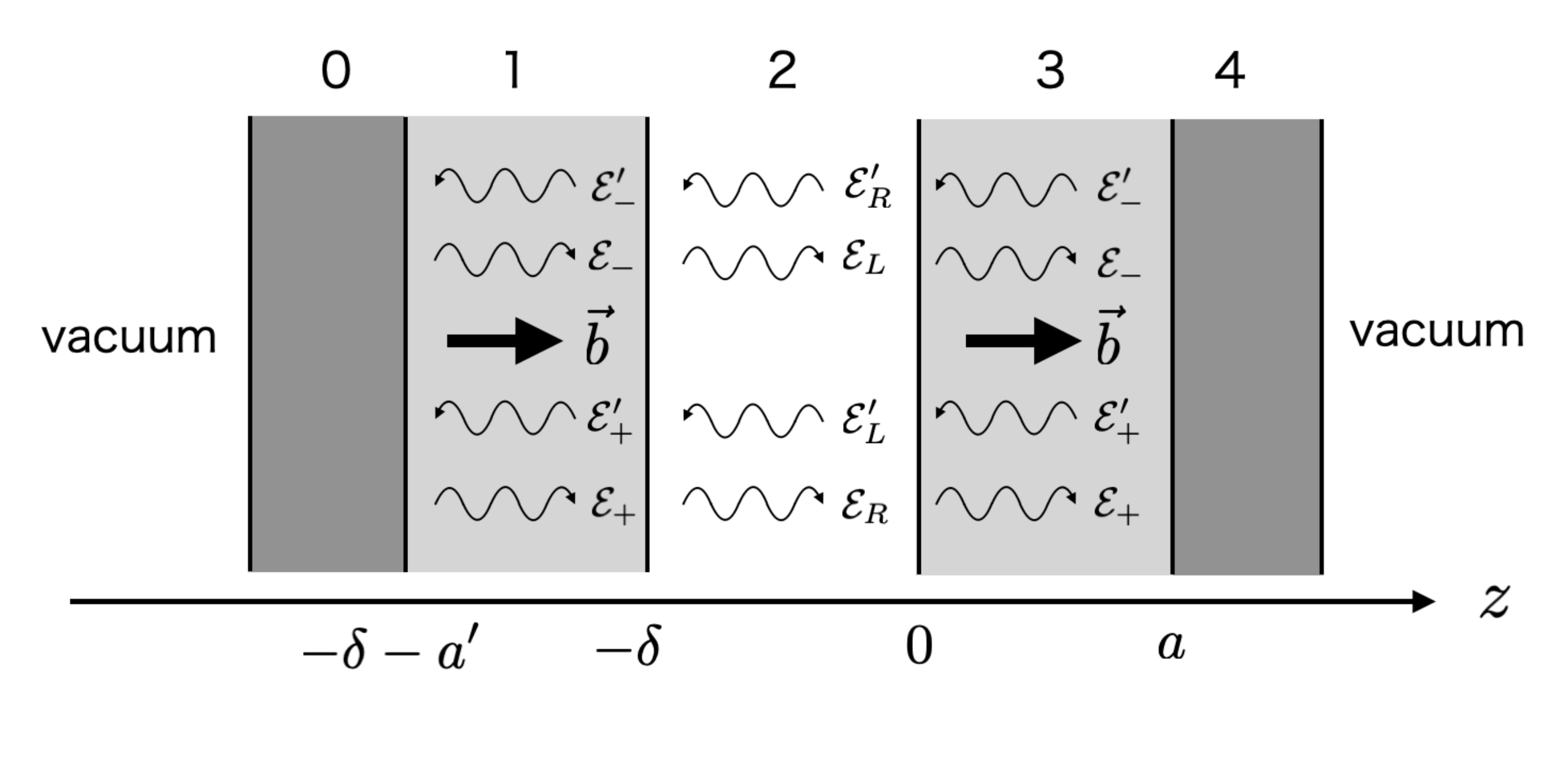}
  \end{center}
  \caption{A schematic picture of the setup 2. Layers 0 and 4 are perfect metals, 1 and 3 are Weyl semimetals with the width $a'$ and $a$ respectively, and layer 2 is the vacuum. Possible eigenmodes are shown. The upper one, which we call the ``$-$'' mode, and the lower one, the ``$+$'' mode, are independent solutions.}
  \label{fig:PWVWP}
\end{figure}

Next let us consider the setup shown in Fig.~\ref{fig:PWVWP}. This setup will be further discussed in the next section 
in connection to new force searches. Again we note that the following calculations are independent of the width of metal plates. 
Possible eigenmodes are schematically shown in the figure. The upper one, which we call ``$-$'' mode, and the lower one, ``$+$'' mode, are independent solutions as far as the outermost bodies are the perfect metal. Otherwise, they are mixed. In this paper, we just focus on the case of thick perfect metals as the layer 0 and 4.
One solution (which we call ``+'' mode) is of the form:
\begin{align}
	&\vec E =\begin{cases}
	 \mathcal E_{1+} e^{i k_z^+ z}\hat e_+ + \mathcal E'_{1+} e^{-i k_z^+ z} \hat e'_+ & {\rm for~}-\delta-a' < z < -\delta \\
	  \mathcal E_{2R} e^{i q_z z}\hat e_R + \mathcal E'_{2L} e^{-i q_z z} \hat e'_L & {\rm for~}-\delta < z < 0 \\
	  \mathcal E_{3+} e^{i k_z^+ z}\hat e_+ + \mathcal E'_{3+} e^{-i k_z^+ z} \hat e'_+ & {\rm for~~} 0< z < a \\
	 \end{cases},
	 \\
	&\vec B =\begin{cases}
	\frac{k_+}{\omega}\left(\mathcal E_{1+} e^{i k_z^+ z}\hat e_+^{(B)} + \mathcal E'_{1+} e^{-i k_z^+ z} \hat e_+^{'(B)}\right) & {\rm for~}-\delta-a' < z < -\delta \\
	\frac{q}{\omega}\left(\mathcal E_{2R} e^{i q_z z}\hat e_R^{(B)} + \mathcal E'_{2L} e^{-i q_z z} \hat e_L^{'(B)}\right) & {\rm for~}-\delta < z < 0 \\
	\frac{k_+}{\omega}\left(\mathcal E_{3+} e^{i k_z^+ z}\hat e_+^{(B)} + \mathcal E'_{3+} e^{-i k_z^+ z} \hat e_+^{'(B)}\right) & {\rm for~~} 0< z < a \\
	\end{cases}.
\end{align}
The boundary conditions at the perfect metal plate $z=a$ and $z=-\delta-a'$ give
\begin{align}
	\frac{\mathcal E'_{3+}}{\mathcal E_{3+}} = e^{2ik_z^+a},~~~~~~\frac{\mathcal E'_{1+}}{\mathcal E_{1+}} = e^{-2ik_z^+(\delta+a')}.
\end{align}
The boundary conditions at the surface of the Weyl semimetal $z=0$ and $z=-\delta$ give
\begin{align}
	\frac{\mathcal E_{2R}-\mathcal E'_{2L}}{\mathcal E_{2R}+\mathcal E'_{2L}} = \frac{q_z}{k_z^+} \frac{1-e^{2ik_z^+a}}{1+e^{2ik_z^+a}},~~~~~~
	\frac{\mathcal E_{2R}-\mathcal E'_{2L}e^{2iq_z\delta}}{\mathcal E_{2R}+\mathcal E'_{2L}e^{2iq_z\delta}} = \frac{q_z}{k_z^+} \frac{1-e^{-2ik_z^+a'}}{1+e^{-2ik_z^+a'}},
\end{align}
After some calculations, we eventually find an equation that gives allowed modes in the system:
\begin{align}
	0= 1 - e^{2iq_z\delta} \frac{R_+ + e^{2i k_{z}^+ a}}{1+ R_+ e^{2i k_{z}^+ a}} \frac{R_+ + e^{2i k_{z}^+ a'}}{1+ R_+ e^{2i k_{z}^+ a'}},
\end{align}
where $R_+$ is a reflection coefficient that has been defined in Eq.~(\ref{Rplus}).
We obtain a similar solution for the ``$-$'' mode solution. Thus the function $f_\lambda$ is given as follows:
\begin{align}
	f_\lambda(\omega) = \left[ 1- r_\lambda^{(24)}(\omega) r_\lambda^{(20)} (\omega)\,e^{2iq_z\delta} \right]
	\left( 1+ R_\lambda e^{2i k_{z}^\lambda a} \right) \left( 1+ R_\lambda e^{2i k_{z}^\lambda a'} \right),
	\label{f_PWVWP}
\end{align}
where $\lambda= +$ or $-$ and $r_\lambda^{(ij)}$ denotes the effective reflection coefficient between body $i$ and $j$:\footnote{
	These results depend on the relative direction of $\vec b$ in the two Weyl semimetals. For example, if the direction of $\vec b$ in the body 1 were reversed, Eq.~(\ref{f_PWVWP}) should be modified as $f_{\pm} = \left[1- r_\pm^{(24)} r_\mp^{(20)}\,e^{2iq_z\delta}\right]\left( 1+ R_\pm^{2i k_{z}^\pm a} \right) \left( 1+ R_\mp e^{2i k_{z}^\mp a'} \right)$ . 
}
\begin{align}
	r_\lambda^{(24)}(\omega) =  \frac{R_\lambda + e^{2i k_{z}^\lambda a}}{1+ R_\lambda e^{2i k_{z}^\lambda a}},~~~~~~
	r_\lambda^{(20)}(\omega) =  \frac{R_\lambda + e^{2i k_{z}^\lambda a'}}{1+ R_\lambda e^{2i k_{z}^\lambda a'}}.
\end{align}
Thus the Casimir free energy per unit area is given by
\begin{align}
	\mathcal F_{\rm Cas} =T{\sum_{\lambda={+, -}}}{\sum_{\ell\geq 0}}'\int\frac{d^2k_\parallel}{(2\pi)^2} \ln \left[1- r_\lambda^{(24)} (i\xi_\ell)r_\lambda^{(20)}(i\xi_\ell)\,e^{-2\tilde q_{z,\ell} \delta}\right] + (\delta{\rm-independent}),
\end{align}
where $\tilde q_{z,\ell} = \sqrt{\xi_\ell^2+ k_\parallel^2}$. The Casimir force is given by
\begin{align}
	F_{\rm Cas} &= -\frac{\partial \mathcal F_{\rm Cas}}{\partial \delta} \nonumber \\
	&=-2T\sum_{\lambda={+, -}}{\sum_{\ell\geq 0}}' \int \frac{d^2 k_\parallel}{(2\pi)^2}\frac{\tilde q_{z,\ell} \,r_\lambda^{(24)} r_\lambda^{(20)} e^{-2\tilde q_{z,\ell} \delta}}{1-  r_\lambda^{(24)} r_\lambda^{(20)} e^{-2\tilde q_{z,\ell} \delta} }.
	\label{Fcas_PWVWP}
\end{align}

\begin{figure}[t]
\begin{center}
   \includegraphics[width=8cm]{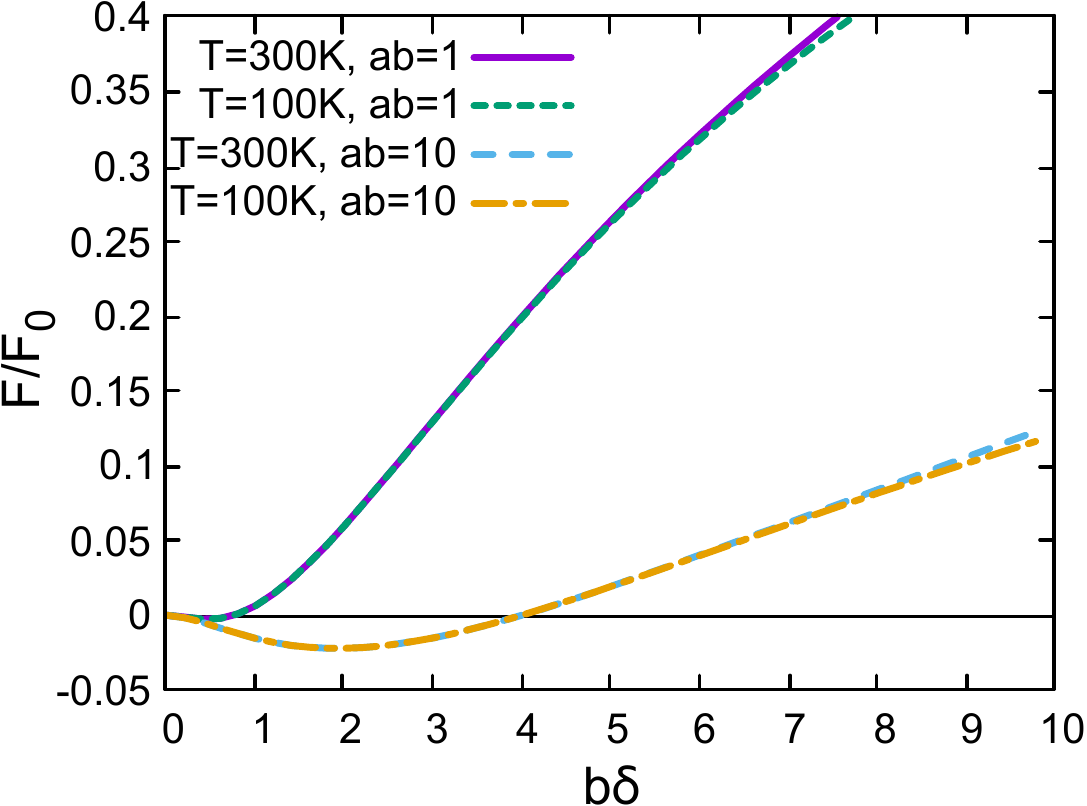}
   \includegraphics[width=8cm]{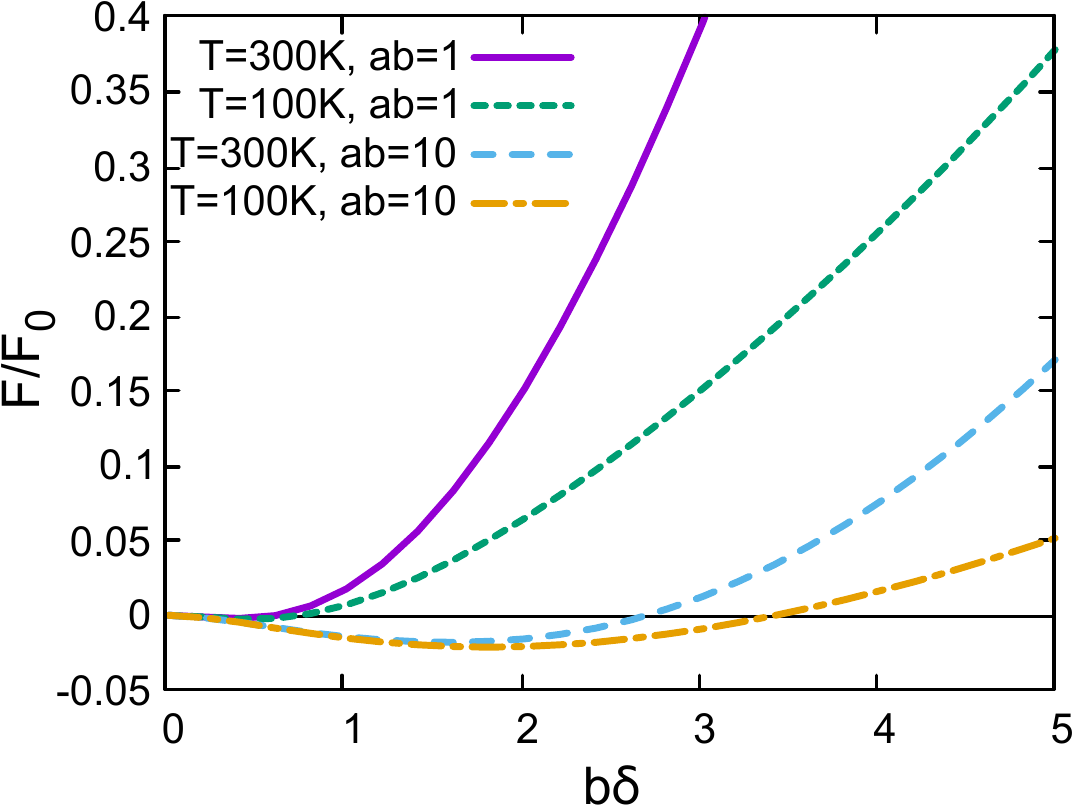}
  \end{center}
  \caption{The Casimir force in the setup 2 (Fig.~\ref{fig:PWVWP}). In the left (right) panel we take $b=1\,{\rm eV}$ ($0.1\,{\rm eV}$) and varied the temperature as $T=300$\,K and $100$\,K with $ab=1$ and $10$. 
  For reference, $1/b=0.197\,\mu {\rm m}$ in the left panel and $1/b=1.97\,\mu {\rm m}$ in the right panel. The negative $F/F_0$, below the black horizontal line, means the repulsive Casimir force.}
  \label{fig:cas_PWVWP}
\end{figure}

Fig.~\ref{fig:cas_PWVWP} shows the numerical result for the Casimir force with $a = a'$, normalized by the standard Casimir force $F_0=-\pi^2/(240 \delta^4)$. In the left (right) panel we take $b=1\,{\rm eV}$ ($0.1\,{\rm eV}$). For reference, $1/b=0.197\,\mu {\rm m}$ in the left panel and $1/b=1.97\,\mu {\rm m}$ in the right panel. 
It is seen that the Casimir force is repulsive for small $\delta$ and the force becomes zero around $b\delta = 4$ for $ab \gg 1$ in the low-temperature limit in both panels. The case without the metals in the zero-temperature limit has been considered in Ref.~\cite{Wilson:2015wsa}, which is consistent with ours in the limit $ab\to \infty$. On the other hand, in the limit $a\to 0$, we obtain $F/F_0=1$ as expected.
Remarkably, the point at which the Casimir force vanishes is stationary, meaning that such a configuration is stable against perturbation. 
This configuration may serve as an ideal setup for the new force search, as explained in the next section.

\section{Implication for new force search}  \label{sec:newforce}

If there exists a light new particle, it generally mediates new force between bodies~\cite{Adelberger:2003zx,Dobrescu:2006au,Damour:2010rp,Fichet:2017bng,Brax:2017xho,Berge:2017ovy,Costantino:2019ixl}. 
In new physics models beyond the Standard Model there often appear light new particles. Examples of such new particles are CP-even scalars, such as dilaton or moduli, axion or axion-like particles or gauge bosons in association with new gauge symmetry. They are also well-motivated dark matter candidates. 
Light CP-even scalar particles may appear in extra dimension theories such as string theory, and they couple to the mass of the material. 
The $B-L$ vector boson may appear in a gauged U(1)$_{B-L}$ extension of the Standard Model, which may be motivated by the explanation of neutrino masses or the grand unified theory. 
The $B-L$ vector boson couples to the $B-L$ number of the material, which is equivalent to the number of neutrons in an electrically neutral body.
In both cases, a potential between small bodies with masses $m_1$ and $m_2$ is given by
\begin{align}
	V = - \frac{Gm_1 m_2}{r}\left(1+\alpha e^{-r/\lambda}\right),  \label{Vpoint}
\end{align}
where $G$ is the Newton constant and $\lambda$ is the Compton wavelength of the new particle. The first term describes the standard Newtonian gravity force and the second term is the new force mediated by the new particle, where $\alpha$ parametrizes the coupling between the new particle and the Standard Model particles.
Even for the CP-odd particles like the axion, a similar form of new force may appear at the loop level~\cite{Costantino:2019ixl,Ferrer:1998ue}.
Various experiments give constraints on the new force on various length scales, corresponding to various mass ranges 
of the new particle. 
The Casimir force measurement gives the most stringent constraint on the new force mediated by a new particle with a mass of $\mathcal O({\rm eV})$ at the separation scale of $\mathcal O(0.1$--$1)\,{\rm \mu m}$~\cite{Decca:2005qz,Decca:2007yb,Sushkov:2011md,Chen:2014oda}.

From the viewpoint of new force search, the Casimir force is an obstacle that hides the new force effect (see e.g., Ref.~\cite{Bennett:2018ylp} for an idea of shielding the Casimir-Polder force for new force search). 
Our setup described in Sec.~\ref{sec:PWVWP} opens up a possibility of the Casimir-free stationary point at the separation of $\mathcal O(0.1$--$1)\,{\rm \mu m}$ scales, and hence the sensitivity to the new force may be significantly improved.
Actually, uncertainties of the theoretical calculation of the Casimir force give a large amount of systematic error, which limits the sensitivity to the new force~\cite{Decca:2007yb}. In our setup discussed below, we can set the metal/Weyl-semimetals to the stationary point in which the Casimir force vanishes and hence the theoretical uncertainties for the Casimir force calculation are irrelevant whatever the origin of the uncertainty is. Therefore, in principle, the sensitivity will be improved just by reducing the statistical/random error.

\begin{figure}[t]
\begin{center}
   \includegraphics[width=16cm]{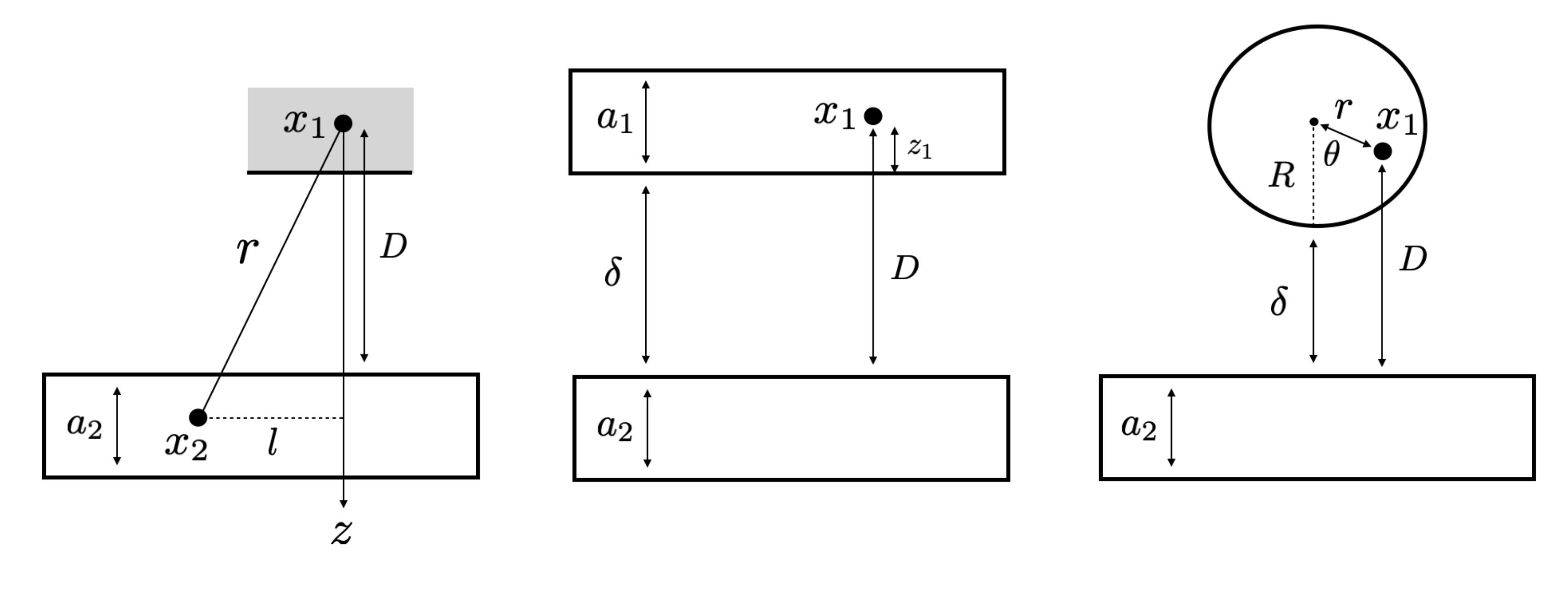}
  \end{center}
  \caption{(Left) Coordinate to calculate the potential between $x_1$ in the body 1 and $x_2$ in the body 2. (Middle) The case where body 1 is a plate. (Right) The case where body 1 is a sphere.}
  \label{fig:new}
\end{figure}

Let us derive a new force acting on 2 bodies. 
For that purpose, we first calculate the potential induced by an exchange of a new particle between a small piece of body 1 with mass $m_1$ and the entire body 2 (see the left panel of Fig.~\ref{fig:new} for the choice of coordinate):
\begin{align}
	dV &=  -G\alpha m_1 \rho_2 \int \frac{e^{-r/\lambda}}{r} d^3 x_2 \nonumber \\
	&= -G\alpha m_1 \rho_2 \int_{D}^{D + a_2} dz \int_0^{L} 2\pi l  \frac{e^{-\sqrt{z^2 + l^2}/\lambda}}{\sqrt{z^2 + l^2}}dl \nonumber \\
	&= -2\pi G\alpha m_1 \rho_2 \lambda^2 e^{-D/\lambda} \left(1-e^{-a_2/\lambda} \right),
	\label{dV12}
\end{align}
where we have made an approximation $\lambda \ll L$ and $\rho_1$ and $\rho_2$ are the mass density of the body 1 and 2, respectively.\footnote{
	In the case of the $B-L$ vector boson, $\rho_1$ and $\rho_2$ should be regarded as the mass density times the neutron number fraction in each atom, which is typically $\sim 0.5$.
}
The total potential between body 1 and 2 is obtained by integrating $dV$ over body 1. Below we consider two cases as the shape of the body 1: a parallel plate or a sphere.

If body 1 is a parallel plate, as shown in the middle panel of Fig.~\ref{fig:new}, the total potential is given by the integration of $dV$ with $d^3x_1$. By noting $m_1=\rho_1d^3x_1$ and the integration range is expressed as $\int d^3x_1 e^{-D/\lambda} = \int_0^{a_1} dz_1 \int dS_1 e^{-(\delta + z_1)/\lambda}$, we obtain
\begin{align}
	V(\delta) = - S_1 \times 2\pi G\alpha \rho_1\rho_2   \lambda^3 e^{-\delta/\lambda} \left(1-e^{-a_1/\lambda} \right)\left(1-e^{-a_2/\lambda} \right),
\end{align}
where $S_i$ is the cross sectional area of the body $i$ $(=1,2)$, assuming $S_1 < S_2$. The new force per unit area between bodies 1 and 2 is given by
\begin{align}
	F_{\rm new} = -\frac{1}{S_1}\frac{\partial V}{\partial \delta}
	= -2\pi G\alpha \rho_1\rho_2 \lambda^2 e^{-\delta/\lambda} \left(1-e^{-a_1/\lambda} \right)\left(1-e^{-a_2/\lambda} \right).
	\label{Fnew_plate}
\end{align}

If body 1 is a sphere, as shown in the right panel of Fig.~\ref{fig:new}, the total potential between the two bodies is given by
\begin{align}
	V(\delta) &= \int d^3 x_1 dV \nonumber \\
	&=  - 2\pi G\alpha \rho_1 \rho_2 \lambda^2 e^{-(R+\delta)/\lambda} \left(1-e^{-a_2/\lambda} \right) \left(2\pi \int_0^R dr \int_0^\pi d\theta \, r^2\sin\theta e^{r\cos\theta/\lambda} \right) \nonumber \\
	&= - 4\pi^2 G\alpha \rho_1\rho_2 \lambda^4 R\, e^{-\delta/\lambda} \left(1-e^{-a_2/\lambda} \right),
\end{align}
where we used $D=R+\delta-r\cos\theta$ and assumed $\lambda \ll R$ in the last line.
The total new force between bodies 1 and 2 is given by (note that it is the total force, not the force per unit area)
\begin{align}
	F_{\rm new}=  -\frac{\partial V}{\partial \delta}
	= - 4\pi^2 G\alpha \rho_1\rho_2 \lambda^3 R\, e^{-\delta/\lambda} \left(1-e^{-a_2/\lambda} \right).
\end{align}

\begin{figure}[t]
\begin{center}
   \includegraphics[width=8cm]{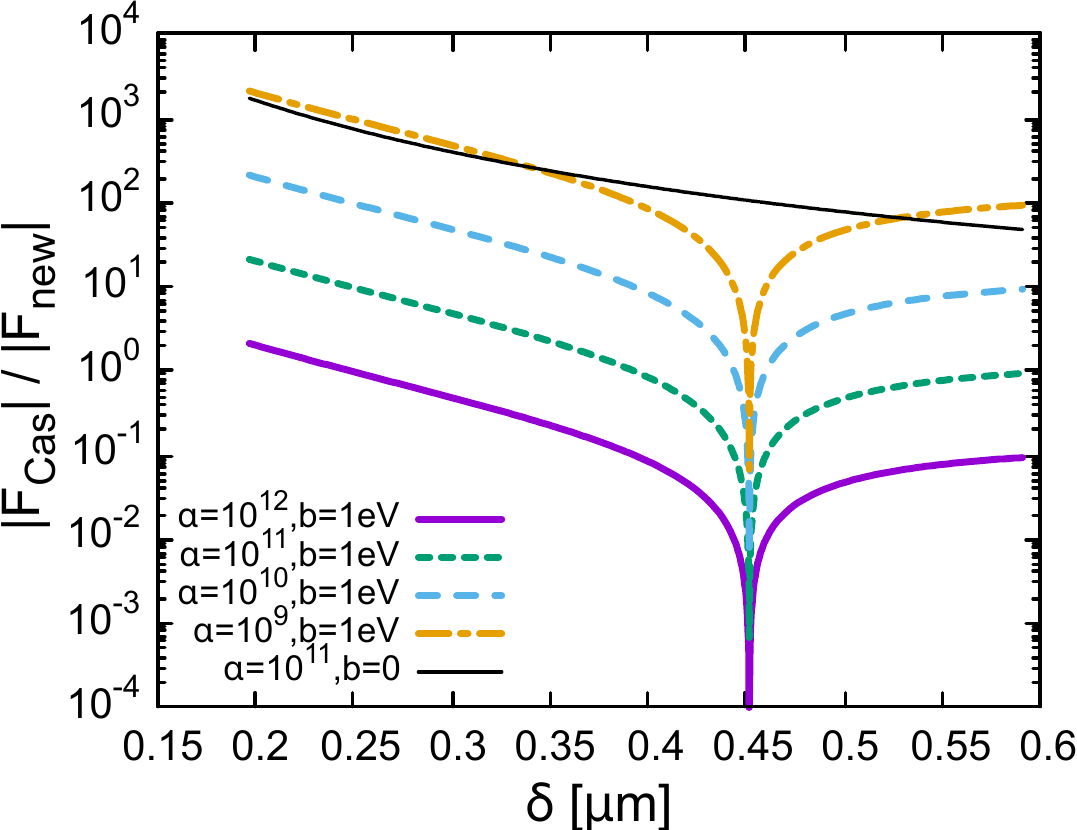}
    \includegraphics[width=8cm]{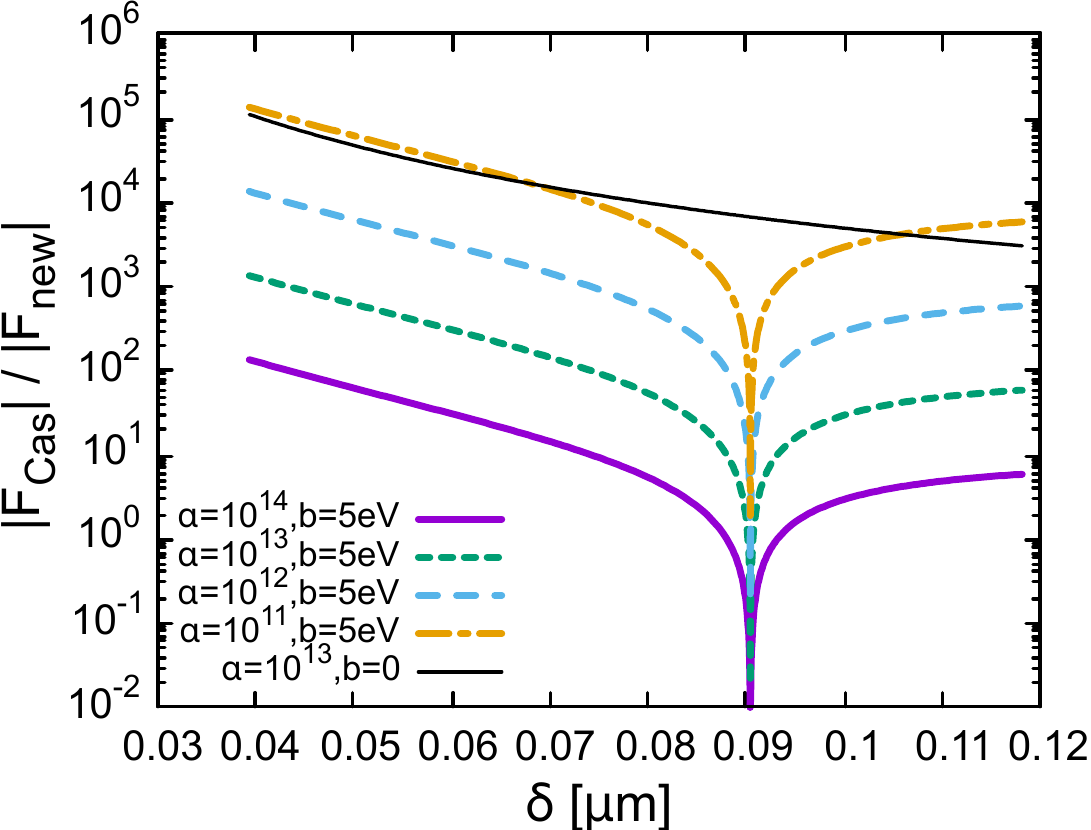}
  \end{center}
  \caption{The ratio between the Casimir force $|F_{\rm Cas}|$ and new force $|F_{\rm new}|$ for various choice of new force strength $\alpha$. We have taken $b=1\,{\rm eV}$ and $\lambda=0.5\,{\mu\rm m}$ in the left panel and $b=5\,{\rm eV}$ and $\lambda=0.1\,{\mu\rm m}$ in the right panel. }
  \label{fig:Fratio}
\end{figure}

Now let us compare the Casimir force and the new force. We assume the setup of Fig.~\ref{fig:PWVWP} with $a=a'$. The Casimir force has already been calculated in (\ref{Fcas_PWVWP}). The new force is given by generalization of Eq.~(\ref{Fnew_plate}) to include the layers of Weyl semimetals and metals.
Assuming the gold layer as the outermost bodies, we obtain
\begin{align}
	F_{\rm new} = -2\pi G \alpha \lambda^2 e^{-\delta/\lambda} 
	\left[ \rho_{\rm WS}(1-e^{-a/\lambda}) + \rho_{\rm Au}e^{-a/\lambda} (1-e^{-d_{\rm Au}/\lambda})\right]^2,
\end{align}
where $\rho_{\rm WS}$ and $\rho_{\rm Au}$ are the mass density of the Weyl semimetal and gold, respectively, and $d_{\rm Au}$ is the width of the gold.
Fig.~\ref{fig:Fratio} shows the ratio between the Casimir force $|F_{\rm Cas}|$ and new force $|F_{\rm new}|$. We have taken $b=1\,{\rm eV}$ and $\lambda=0.5\,{\mu\rm m}$ in the left panel and $b=5\,{\rm eV}$ and $\lambda=0.1\,{\mu\rm m}$ in the right panel. 
The other parameters are taken as $a=2/b$, $T=300\,{\rm K}$ (although the temperature dependence is not significant),
$\rho_{\rm Au}=19.3\,{\rm g/cm^3}$, $d_{\rm Au}=0.2\,{\mu\rm m}$ and $\rho_{\rm WS}=10\,{\rm g/cm^3}$. The new force strength, $\alpha$, is varied as indicated in each panel. For reference, we also plotted the case of $b=0$.
It is clearly seen that the Casimir force is suppressed by orders of magnitude compared with the case of $b=0$. Note that the current upper bound on $\alpha$ is about $\alpha\sim 5\times 10^{10}$ for $\lambda=0.5\,{\mu\rm m}$ and $\alpha\sim 10^{13}$ for $\lambda=0.1\,{\mu\rm m}$~\cite{Decca:2005qz,Sushkov:2011md}.
Even for $\alpha$ much below this upper bound, the Casimir force can be negligible and the new force can be the dominant force acting on the body at some certain distance, which opens up a possibility to improve the new force search.
We also note that the distance at which the Casimir force vanishes depends on $b$, $a$ and temperature, as seen from Fig.~\ref{fig:cas_PWVWP} and hence it may be possible to adjust it to the desired point.
However, our discussion here is just a short demonstration of our idea and its theoretical background, and careful studies are required for real experiments.

\begin{figure}[t]
\begin{center}
   \includegraphics[width=8cm]{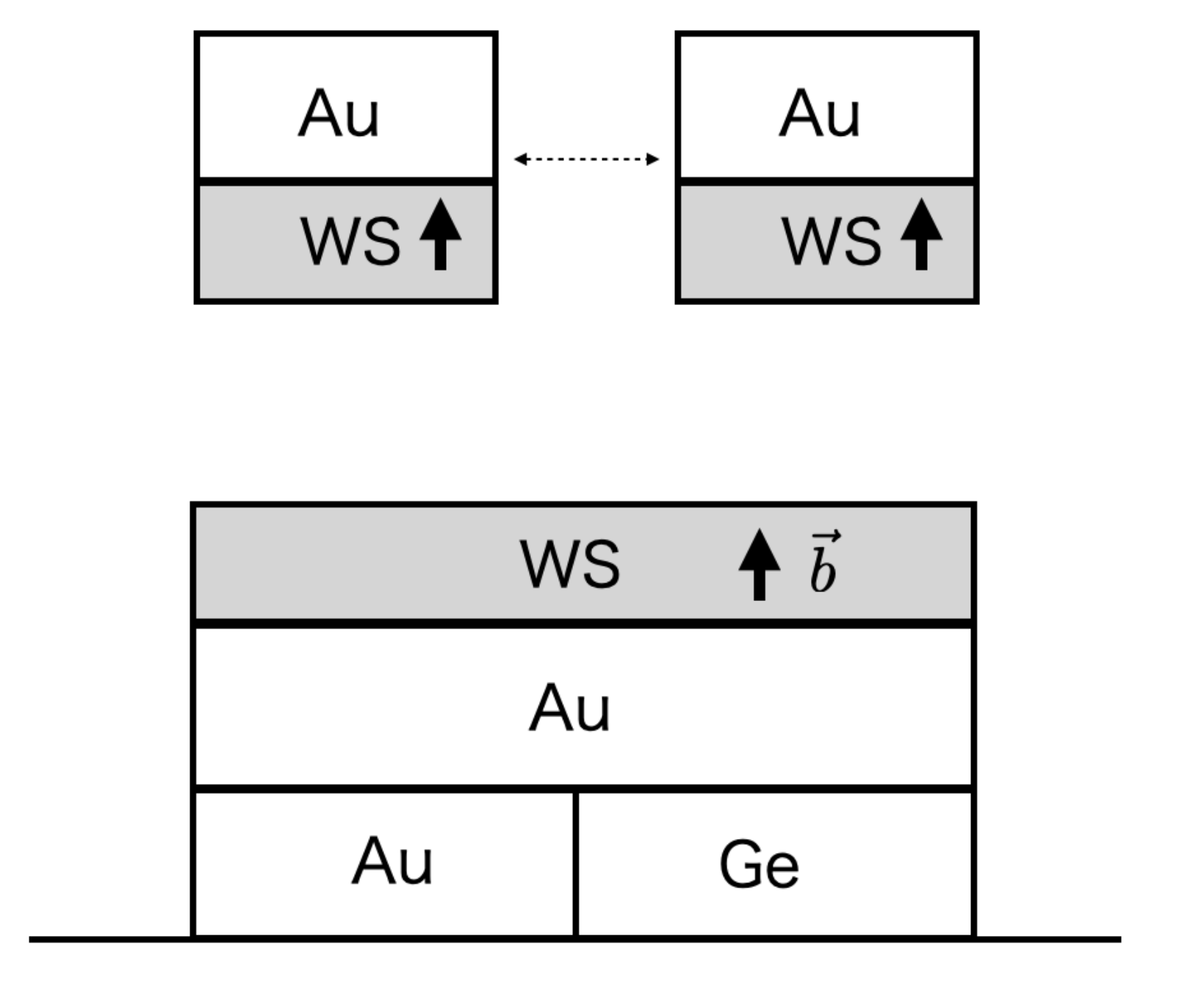}
  \end{center}
  \caption{Schematic picture of the setup to measure the difference of the stationary point between two materials, gold (Au) and germanium (Ge). WS denotes the Weyl semimetal with the arrow indicating the direction of $\vec b$. A thin dotted arrow indicates that the upper body is moved in a  parallel direction to take a difference. }
  \label{fig:exp}
\end{figure}

As a final remark, the sensitivity may be further improved by taking a difference of force between different experimental setups along the line of Ref.~\cite{Decca:2005qz}, as schematically shown in Fig.~\ref{fig:exp}. 
Suppose that we found a stationary point in an experiment at which the Casimir force vanishes and it can be compared with the theoretical prediction.
However, taking uncertainties in the calculation into account, it may be rather difficult to extract a contribution from the new force if it is sufficiently weak. 
Still one can take a difference of the stationary point between two cases, e.g., the case of gold and germanium as the outermost body as shown in Fig.~\ref{fig:exp}. 
If the common gold layer of the body is thick enough, i.e., thicker than the plasma wavelength $\sim 0.135\,{\rm {\mu}m}$, the Casimir force acting on the body is the same between the case of gold and germanium since the Casimir force becomes independent of what lies beyond the thick gold layer, while the new force ``feels'' the materials beyond the coating layer since the mass density is different.
Thus the difference should be zero if there were no new force, and finding a nonzero difference would be a signal of a new force.


\section{Conclusions and discussion} \label{sec:con}

Since the theoretical proposal long ago, the Casimir force had been measured in various setups. It is typically an attractive force, but a repulsive force is also possible in some setups. 
The measurement of the Casimir force itself is interesting since it gives direct evidence of the quantum nature of the vacuum. On the other hand, the Casimir force is a major obstacle to find a new force mediated by new hypothetical particles at the micro-meter scale.
We point out that it is possible to achieve zero Casimir force in experimental setups involving Weyl semimetals.
Since we can set the metals/Weyl-semimetals at the stationary position where the Casimir force vanishes, we do not need to worry about uncertainties for the theoretical calculation of the Casimir force, and hence this setup may potentially enhance the sensitivity of the new force search by just reducing experimental errors.

Although we have considered Weyl semimetals at room temperature as an example, other nonreciprocal materials can be used, such as indium antimonide (InSb), where an external magnetic field needs to be applied~\cite{Fuchs:2017}. 
In Refs.~\cite{Grushin:2010,Rodriguez-Lopez:2013pza}, Casimir force on topological or Chern insulators has been calculated and a repulsive force is found.
Without nonreciprocal materials, nonequilibrium systems consisting of reciprocal materials allow zero Casimir force~\cite{Kruger:2012,Messina:2011,Iizuka:2017,Chen:2016}. In Ref.~\cite{Iizuka:2021}, the nonequilibrium Casimir force in a system consisting of SiO$_2$ and silicon plates was made stable around the zero-force point through a feedback control process involving temperature measurement and modulation.
There are several possibilities for obtaining Casimir force-free setups.
We will study which setups are the most appropriate in actual experiments in future works\footnote{Note that we have found a stabilizing point for the force, not merely a zero force. This ``Casimir spring" may have the potential for various applications beyond fundamental physics.}.

\paragraph{Note added}
While finalizing this manuscript, Ref.~\cite{Favitta:2023hlx} appeared on arXiv. They also discuss the finite-temperature effect on the Casimir force in axion electrodynamcis.

\section*{Acknowledgment}

This work was supported by JSPS KAKENHI Grant Nos.\ 17H06359 (K.N.), 18K03609 (K.N.).
This work was supported by World Premier International Research Center Initiative (WPI), MEXT, Japan.  
Y.E.\, is supported in part by the DOE grant DE-SC0011842.
K.M.\, was supported by MEXT Leading Initiative for Excellent Young Researchers Grant No.\ JPMXS0320200430,
and by JSPS KAKENHI Grant No.\ 	JP22K14044.

\appendix

\section{Lifshitz formula for Casimir force at finite temperature}  \label{app:Lif}

\subsection{Lifshitz formula}

For given setups including metals or dielectrics, the frequency of electromagnetic waves are generally discretized.
To derive the Casimir energy, we should evaluate the discrete summation over the frequency $\omega_{n,\lambda}$ 
\begin{align}
	\sum_{\lambda}{\sum_{n\geq 0}}' \omega_{n,\lambda},
\end{align}
where $\lambda$ represents different types of modes for a given setup and the prime in the summation implies that the $n=0$ term should be multiplied by a factor $1/2$. For systems consisting of normal metals or dielectrics, a common choice is that we classify the modes into the transverse-magnetic (TM) and transverse-electric (TE) modes. In this case, $\lambda =$ TM or TE. Another choice is to classify it into right- and left-handed polarized waves, as done in the main text of this paper, since it is more convenient in the case of chiral media such as Weyl semimetals. In this case, $\lambda =$ R or L.\footnote{
	In the main text, we assumed a perfect mirror at the boundary of the system. If we take into account the effect of the finite dielectric constant of the metal, left and right modes are mixed and the classification by the index $\lambda$ may no longer be convenient. 
}
It is conveniently evaluated by using the argument principle:
\begin{align}
	\sum_{\lambda}{\sum_{n\geq 0}}' \omega_{n,\lambda} =\sum_{\lambda} \frac{1}{2\pi i}\oint \omega \,d\ln f_\lambda (\omega).
	\label{omegaflambda}
\end{align}
Here the function $f_\lambda(\omega)$ is chosen such that the solution to $f_\lambda(\omega)=0$ gives possible allowed frequencies $\omega$ satisfying the boundary condition in the system under consideration and also that $1/f_\lambda(\omega)$ contains first-order poles at $\omega=\omega_{n,\lambda}$.

In the finite temperature system with temperature $T$, the Casimir free energy density per unit area is given by
\begin{align}
	\mathcal F_{\rm Cas}= \sum_{\lambda} {\sum_{n\geq 0}}' \int \frac{d^2k_\parallel}{(2\pi)^2} \left[ \frac{\omega_{n,\lambda}}{2} + T\ln\left(1-e^{-\omega_{n,\lambda}/T}\right)\right] - \mathcal F_{\rm vac},
\end{align}
where $\mathcal F_{\rm vac}$ denotes the vacuum contribution to regularize the divergence coming from the quantum zero-point energy. By the generalized argument principle, it is further conveniently evaluated as
\begin{align}
	\mathcal F_{\rm Cas} &= T{\sum_{\lambda}}\int\frac{d^2k_\parallel}{(2\pi)^2} {\sum_{n\geq 0}}' \ln\left(2\sinh\frac{\omega_{n,\lambda}}{2T}\right) 
    \nonumber \\
	&= T{\sum_{\lambda}}\int\frac{d^2k_\parallel}{(2\pi)^2} \frac{1}{2\pi i}\oint \ln\left(2\sinh\frac{\omega_{\lambda}}{2T}\right) d\ln f_\lambda(\omega)
    \nonumber \\
	&=T{\sum_{\lambda}}{\sum_{\ell\geq 0}}'\int\frac{d^2k_\parallel}{(2\pi)^2} \ln f_\lambda(i \xi_\ell),
	\label{CasLif}
\end{align}
with the function $f_\lambda$ where $\xi_\ell=2\pi T \ell$ denotes the Matsubara frequency. After all, we have replaced the summation over $n$ with the summation over $\ell$. The latter formulation is much more convenient for practical purposes for general materials with finite dielectric constants, since we do not need to explicitly solve the complicated equation $f_\lambda(\omega)=0$ in the latter formulation. 
Note that we have regularized the divergence with the counter term $\mathcal F_{\rm vac}$.
This is a general formula and the information of the concrete setup is contained in the form of the function $f_\lambda(\omega)$. The expression for the zero temperature limit is obtained by formally replacing
\begin{align}
	T{\sum_{\ell\geq 0}}'  \to \int_0^\infty\frac{d\xi}{2\pi}.
\end{align}
Below we evaluate the Casimir force for the three-layer and five-layer cases, by using the concrete function $f_\lambda(\omega)$ obtained in the previous section.


\subsection{An example of normal metal}

\begin{figure}[t]
\begin{center}
   \includegraphics[width=12cm]{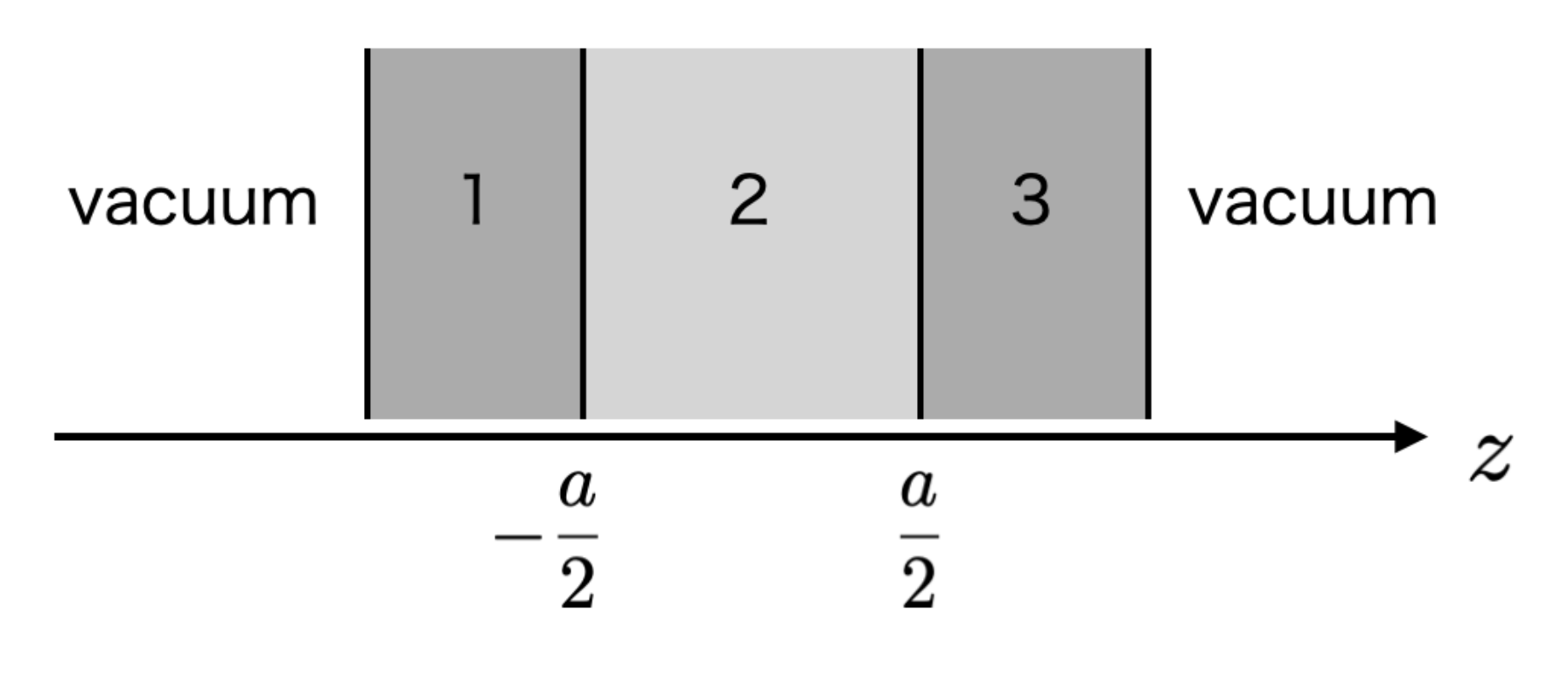}
  \end{center}
  \caption{The layers 1, 2, and 3 are assumed to have dielectric constants of $\epsilon_1, \epsilon_2$ and $\epsilon_3$, respectively. The leftmost and rightmost regions are assumed to be the vacuum.}
  \label{fig:3lay}
\end{figure}

As a simple example, we consider the three-layer setup as shown in Fig.~\ref{fig:3lay}. The dielectric constants in each layer 1,2 and 3 are taken to be $\epsilon_i$ $(i=1,2,3)$. As for the magnetic permeability, we take $\mu_1=\mu_2=\mu_3=1$ for simplicity. We want to know the allowed modes and the dispersion relation of the photon in this setup. 
We assume the solution of the form 
\begin{align}
	\vec E(\vec x,t) = \vec{\mathcal E}(z)e^{i(\vec k_\parallel\cdot\vec x_\parallel-\omega t)},~~~~~~
	\vec B(\vec x,t) = \vec{\mathcal B}(z)e^{i(\vec k_\parallel\cdot\vec x_\parallel-\omega t)}.
\end{align}

First consider the TM mode, i.e., $\mathcal B_z=0$. We seek a solution of the form
\begin{align}
	\mathcal E_z = 
	\begin{cases}
	Ae^{-q_3 z} & z > a/2 \\
	Be^{q_2z} + Ce^{-q_2z} & -a/2< z < a/2 \\
	De^{q_1 z} & z < -a/2
	\end{cases},
\end{align}
with $A, B, C, D$ denoting constants (one of which is redundant because the overall normalization is free) and $k_{\parallel}^2-q_i^2 = \epsilon_i\omega^2$. 
Note that we assume $q_3$ and $q_1$ are real and positive so that the electromagnetic field damps at large $|z|$, while $q_2$ can be either real or pure imaginary. 
The boundary condition implies continuous $\epsilon\mathcal E_z$ across the boundary. Also, by noting that $\vec \nabla\cdot \vec E = 0$ for all the regions (since we are considering the dielectric material), we have $i\vec k_{\parallel}\cdot \vec E_\parallel = -\partial_z E_z$. Since $\vec E_\parallel$ must be continuous across the boundary, this implies that $\partial_z \mathcal E_z$ is continuous. To summarize, the boundary condition at $z=a/2$ is
\begin{align}
	\epsilon_2 \mathcal E_z(z\to a/2-0) = \epsilon_3 \mathcal E_z(z\to a/2+0),~~~~~~
	\partial_z \mathcal E_z(z\to a/2-0) = \partial_z \mathcal E_z(z\to a/2+0).
\end{align}
From this we obtain
\begin{align}
	\frac{B}{C} = \frac{\epsilon_3 q_2 - \epsilon_2 q_3}{\epsilon_3 q_2 + \epsilon_2 q_3} e^{-q_2 a} = r^{(23)}_{\rm TM}  e^{-q_2 a}.
\end{align}
Here we have defined the reflection coefficients for the TM and TE mode as~\cite{Jackson:1998nia}
\begin{align}
	r_{\rm TM}^{(ij)}(\omega) = \frac{\epsilon_jq_i - \epsilon_i q_j}{\epsilon_jq_i + \epsilon_i q_j}, ~~~~~~r_{\rm TE}^{(ij)}(\omega) = \frac{q_i - q_j}{q_i + q_j},
\end{align}
where $q_j^2(\omega) = k_\parallel^2 - \epsilon_j(\omega) \omega^2$.
Similarly, from the boundary condition at $z=-a/2$, we obtain
\begin{align}
	\frac{C}{B} = r^{(21)}_{\rm TM}  e^{-q_2 a}.
\end{align}
Combining these boundary conditions, we have 
\begin{align}
	1 = e^{-2q_2a}  r^{(21)}_{\rm TM}  r^{(23)}_{\rm TM}.   \label{TMdis}
\end{align}
This determines the allowed value of $\omega$. Since $\epsilon$ are generally complicated functions of $\omega$, this is a highly nonlinear equation in general.
In the limit of perfect metal for the body 1 and 3, $\epsilon_1=\epsilon_3 \to -\infty$ (note that $q_{1,3} \simeq \sqrt{-\epsilon_{1,3}}\omega$), the solution is given by $q_2= i n\pi/a$ $(n=0,\pm1,\dots)$. 

Next consider the TE mode, i.e., $\mathcal E_z=0$. Similarly, we obtain
\begin{align}
	1 = e^{-2q_2a}  r^{(21)}_{\rm TE}  r^{(23)}_{\rm TE}.   \label{TEdis}
\end{align}
In the limit of perfect metal, the solution is given by $q_2= i n\pi/a$ $(n=0,\pm1, \dots)$. 

Conditions (\ref{TMdis}) and (\ref{TEdis}) are expressed in a combined form as, 
\begin{align}
	0 = f_\lambda(\omega) \equiv 1- r^{(21)}_\lambda(\omega)r^{(23)}_\lambda(\omega) e^{-2q_2a},~~~~~~q_2=\sqrt{k_\parallel^2-\epsilon_2\omega^2}.
	\label{flambda_3}
\end{align}
with $\lambda={\rm TM, TE}$. This function $f_\lambda(\omega)$ will be used for the evaluation of Casimir force in this system.

Now the function $f_\lambda(\omega)$ is determined, and the Casimir free energy density is given by 
\begin{align}
	\mathcal F_{\rm Cas}=T\sum_{\lambda = {\rm TE, TM}}{\sum_{\ell\geq 0}}' \int \frac{d^2 k_\parallel}{(2\pi)^2}\ln\left[1- r^{(21)}_\lambda(i\xi_\ell) r^{(23)}_\lambda(i\xi_\ell) e^{-2aq_\ell}\right],
\end{align}
where $q_\ell = \sqrt{k_\parallel^2+\epsilon_2(i\xi_\ell)\xi_\ell^2}$. The Casimir force is given by
\begin{align}
	F_{\rm Cas} =-\frac{\partial \mathcal F_{\rm Cas}}{\partial a}
	=-2T\sum_{\lambda = {\rm TE, TM}}{\sum_{\ell\geq 0}}' \int \frac{d^2 k_\parallel}{(2\pi)^2}\left[\frac{r^{(21)}_\lambda(i\xi_\ell) r^{(23)}_\lambda(i\xi_\ell) e^{-2aq_\ell} q_{\ell}}{1- r^{(21)}_\lambda(i\xi_\ell) r^{(23)}_\lambda(i\xi_\ell) e^{-2aq_\ell}}\right].
	\label{F_3layer}
\end{align}
This is the Lifshitz formula~\cite{Lifshitz1956,dzyaloshinskii1961} for the Casimir force at finite temperature with general dielectric constants. We can also obtain the formula for the case of multiple layers~\cite {Tomas:1995,Tomas:2002swv,Raabe:2002}.



\end{document}